\documentstyle[12pt,epsfig]{article} 
\def\baselinestretch{1.3}
\newcommand{\ba}{\begin{array}}
\newcommand{\ea}{\end{array}}
\newcommand{\bd}{\begin{displaymath}}
\newcommand{\ed}{\end{displaymath}}
\newcommand{\be}{\begin{equation}}
\newcommand{\ee}{\end{equation}}
\newcommand{\bea}{\begin{eqnarray}}
\newcommand{\eea}{\end{eqnarray}}
\newcommand{\bei}{\begin{itemize}}
\newcommand{\eei}{\end{itemize}}

\def\a{\alpha}

\def\b{\beta}

\def\l{\lambda}

\def\q2 {q^2}

\def\bt{\begin{table}}
\def\et{\end{table}}

\catcode`@=11 
\def \gsim{\mathrel{\mathpalette\@versim>}}
\def \lsim{\mathrel{\mathpalette\@versim<}}
\def \@versim#1#2{\lower0.4ex\vbox{\baselineskip\z@skip\lineskip\z@skip
     \lineskiplimit\z@\ialign{$\m@th#1\hfil##\hfil$%
     \crcr#2\crcr\sim\crcr}}}
\catcode`@=12 

\begin{document}

\begin{flushright}
{  HRI-P-09-03-001\\
RECAPP-HRI-09-010}
\end{flushright}

\begin{center}

{\large \bf Gaugino mass non-universality in an SO(10) supersymmetric
Grand Unified Theory: low-energy spectra and collider signals }\\[15mm]
Subhaditya Bhattacharya\footnote{E-mail: subha@hri.res.in} and
Joydeep Chakrabortty\footnote{E-mail: joydeep@hri.res.in}\\
 {\em Regional Centre for Accelerator-based Particle Physics \\
     Harish-Chandra Research Institute\\
Chhatnag Road, Jhunsi, Allahabad - 211 019, India}
\\[20mm] 
\end{center}

\begin{abstract} 
We derive the non-universal gaugino mass ratios in a supergravity (SUGRA) 
framework where the Higgs superfields belong to the non-singlet 
representations {\bf 54} and {\bf 770} in an $SO(10)$ Grand Unified Theory (GUT).
We evaluate the ratios for the phenomenologically viable intermediate breaking 
chain $SU(4)_C \times SU(2)_L \times SU(2)_R~(G_{422})$. 
After a full calculation of the gaugino mass ratios, noting some 
errors in the earlier calculation for {\bf 54}, we obtain, using the renormalisation
group equations (RGE), interesting low scale phenomenology of such breaking patterns. 
Here, we assume the breaking of the $SO(10)$ GUT group to the intermediate 
gauge group and that to the Standard Model (SM) take place at the GUT scale itself. 
We also study the collider signatures in multilepton channels at the 
Large Hadron Collider (LHC) for some selected benchmark points allowed by 
the cold dark matter relic density constraint provided by the WMAP data and 
compare these results with the minimal supergravity (mSUGRA) framework with 
similar gluino masses indicating their distinguishability in this regard.

\end{abstract}

\vskip 1 true cm

\newpage
\setcounter{footnote}{0}

\def\baselinestretch{1.5}

\section{Introduction} 

With the Large Hadron Collider (LHC) about to be operative shortly, 
the search for physics beyond the standard model 
(SM) has reached a new height of excitement. Low energy (TeV scale) 
supersymmetry (SUSY) has persistently remained one of the leading candidates 
among scenarios beyond the Standard Model (SM), not only because of its 
attractive theoretical framework, but also for the variety of phenomenological 
implications it offers \cite{Book,Sally,Gl,Martin}. The stabilization 
of the electroweak symmetry breaking (EWSB) scale and the possibility 
of having a cold dark-matter (CDM) candidate with conserved  
$R$-parity ($R = (-1)^{(3B + L + 2S)}$) \cite{CDM} are a few elegant 
phenomenological features of SUSY. Side by side, the possibility of paving 
the path towards a Grand Unified Theory (GUT) is one of its most exciting 
theoretical prospects, where one can relate the $SU(3)$, $SU(2)$ and 
$U(1)$ gauge couplings and the corresponding gaugino masses at a high scale
\cite{Martin,GUT1,GUT2}.

The most popular framework of SUSY-breaking is the minimal supergravity 
(mSUGRA) scheme, where SUSY is broken in the 'hidden sector' via gravity 
mediation and as a result, one can parametrise all the SUSY breaking terms 
by a universal gaugino mass ($M_{1/2}$), a universal scalar mass ($m_{0}$), 
a universal trilinear coupling parameter $A_{0}$, the ratio of the vacuum 
expectation values ($vev$) of the two Higgses ($\tan \beta$) and the 
sign of the SUSY-conserving Higgs mass parameter, ($sgn (\mu)$)
\cite{Martin,mSUGRA}. 

However, within the ambit of a SUGRA-inspired GUT scenario itself, 
one might find some deviations from the above-mentioned simplified 
and idealized situations mentioned above. For instance, the gaugino mass 
parameter ($M_{1/2}$) or the common scalar mass parameter ($m_{0}$) 
can become {\em non-universal} at the GUT scale. In this particular work, 
we adhere to a situation with non-universal gaugino masses in a 
supersymmetric scenario embedded in $SO(10)$ GUT group.

Gaugino masses, arising after GUT-breaking and SUSY-breaking at a high scale, 
crucially depend on the gauge kinetic function, as discussed in the next 
section. One achieves universal gaugino masses if the hidden sector fields 
(Higgses, in particular), involved in GUT-breaking, are singlets under the 
underlying GUT group. However if we include the higher dimensional terms 
(dimension five, in particular) in the non-trivial expansion of the 
gauge-kinetic function, the Higgses belonging to the symmetric products 
of the adjoint representation of the underlying GUT group can be 
non-singlets. If these non-singlet Higgses are responsible for GUT breaking, 
the gaugino masses $M_1$, $M_2$ and $M_3$ become non-universal at the high scale itself. 
It is also possible to have more than one non-singlet representations involved 
in GUT breaking, in which case the non-universality arises from a 
linear combination of the effects mentioned above.

Although this issue has been explored in earlier works, particularly in 
context of the $SU(5)$ \cite{Ellis,Drees,Pran}, there had been one known 
effort \cite{Chamoun} to study the same in 
context of the $SO(10)$. In this paper, we calculate the non-universal
gaugino mass ratios for the non-singlet representations {\bf 54} and 
{\bf 770}, based on the results obtained in \cite{Joydeep}, for the 
intermediate gauge group, namely, Pati-Salam gauge group 
${SU(4)}_{c} \times {SU(2)}_L \times {SU(2)}_R ~~(G_{422})$ with conserved 
$D$-parity \cite {D-parity}. In order to understand the low-energy phenomenology
of such high scale breaking patterns, which is indeed essential in context 
of the LHC and dark matter searches, we scan a wide region of parameter space 
using the renormalisation group equations (RGE) and discuss the consequences
in terms of various low-energy constraints. For example, 
we discuss the consistency of such low-energy spectra with the radiative 
electroweak symmetry breaking (REWSB), Landau pole,  
tachyonic masses etc. We also point out the constraints from stau-LSP 
(adhering to a situation with conserved $R$-parity and hence lightest neutralino-
LSP), and flavour constraints like $b\rightarrow s\gamma$ for all possible combination 
of the parameter space points. Here, we assume the breaking of the $SO(10)$ 
to the intermediate gauge group and that to SM takes place at the GUT scale.
To study the collider signatures in context of the LHC, we choose some 
benchmark points (BP) consistent with the cold dark matter \cite{CDM} relic density constraint 
obtained from the WMAP data \cite{WMAPdata}. 
We perform the so-called 'multilepton channel analysis'
 \cite {Datta:1999uh,baer11} in $same-sign ~dilepton$, $opposite-sign ~dilepton$, 
$trilepton$, $inclusive ~four-lepton$ channels associated with $jets$, as well as 
in $hadronically~ quiet~ trilepton$ channel at these benchmark points. We compare our 
results with WMAP allowed points in mSUGRA, tuned at the same gluino masses 
\footnote {We are benchmarking a potentially faking SUGRA scenario in terms of the 
gluino mass, which has a very important role in the final state event rates. The 
corresponding sfermion masses are fixed by requiring the fulfillment of WMAP 
constraints.}.

There has been a lot of effort in discussing various phenomenological aspects 
\cite{nonugmpheno1,nonugmpheno2,Choi:2007ka,24,Huitu}  
of such high scale non-universality and its effect in terms of collider 
signatures \cite{Bt1,Subho,Priyotosh}, our analysis is remarkable in the 
following aspects:
\begin{itemize}
\item Apart from noting some errors in the earlier available calculations of 
non-universal gaugino mass ratio for the representation {\bf 54}, we present the hitherto 
unknown ratio for the representation {\bf 770} for the intermediate breaking chain $G_{422}$.
\item While we discuss the consistency of the low-energy spectra
obtained from such high scale non-universality in a wide region of the parameter space, 
we study the collider aspects as well in some selected BPs in context of the LHC. 
\item In order to   distinguish such non-universal schemes
from the universal one, we compare our results at these chosen BPs with WMAP 
allowed points in mSUGRA tuned at the same gluino masses. 
We identify a remarkable distinction between the two, which  might be 
important in pointing out the departure in `signature-space' 
\cite{Bourjaily:2005ja,Ash,Barger,ATLAS} in context of the 
LHC for different input schemes at the GUT scale.
\end{itemize} 

The paper is organized as follows. In the following section, we
calculate the non-universal gaugino mass ratios. The low-energy 
spectra, their constancy with various constraints and subsequently
the choices of the BPs have been discussed in section 3.
Section 4 contains the strategy of the collider simulation and the numerical 
results obtained. We conclude in section 5.

\vspace{0.4 cm}

\section {Non-universal Gaugino mass ratios for $SO(10)$}

Here we calculate the non-universal gaugino mass ratios for 
non-singlet Higgses belonging to the representations {\bf 54} and {\bf 770} under $SO(10)$ 
SUSY-GUT scenario.
 
We adhere to a situation where all soft SUSY breaking effects arise
via hidden sector interactions in an underlying supergravity (SUGRA) 
framework, specifically, in $SO(10)$ gauge theories with an arbitrary 
chiral matter superfield content coupled to N=1 supergravity.

All gauge and matter terms including gaugino masses in the N=1 supergravity
Lagrangian depend crucially on two fundamental functions of chiral
superfields \cite{Cremmer}: (i) gauge kinetic function $f_{\alpha \beta}(\Phi)$, 
which is an analytic function of the left-chiral superfields $\Phi_{i}$ and 
transforms as a symmetric product of the adjoint representation of the 
underlying gauge group ($\alpha$, $\beta $ being the gauge indices, 
run from 1 to 45 for $SO(10)$ gauge group); 
and (ii) $G(\Phi_{i},\Phi^*_{i})$, a real function of $\Phi_{i}$ and gauge 
singlet, with $G = K + ln|W|$ ($K$ is the K$\ddot a$hler potential and $W$ is the 
superpotential). 

The part of the N=1 supergravity Lagrangian
containing kinetic energy and mass terms for gauginos and gauge bosons 
(including only terms containing the real part of $f(\Phi)$) reads

\bea
e^{-1} {\mathcal{L}}= -\frac{1}{4} Re f_{\alpha \beta}(\phi)(-1/2
\bar{\lambda}^
{\alpha} D\!\!\!\!/ \lambda^{\beta})-\frac{1}{4} Re f_{\a \b}(\phi)F_{\mu
  \nu}^{\a}F^{\beta \mu \nu} \nonumber \\
 +\frac{1}{4} e^{-G/2}
G^{i}((G^{-1})^{j}_{i})[\partial f^*_{\a \b}(\phi^*)/\partial
{\phi^{*j}}]\l^{\a}\l^{\b} + h.c
\label {lag}
\eea  
\noindent
where $G^i = \partial{G}/\partial{\phi_{i}}$ and $(G^{-1})^{i}_j$ is the
inverse matrix of 
${G^j}_i\equiv \partial{G} / {\partial{\phi^{*i}}\partial{\phi_j}}$,
$\l^{\a}$ is the gaugino field, and $\phi$ is the scalar component of the 
chiral superfield $\Phi$, and $F_{\mu \nu}^{\a}$ is defined in unbroken 
$SO(10)$. The $F$-component of $\Phi$ enters the last term to generate 
gaugino masses with a consistent SUSY breaking with non-zero $vev$ of the chosen 
$\tilde{F}$, where 
\bea
\tilde{F}^{j}=\frac{1}{2} e^{-G/2}
[G^{i}((G^{-1})^{j}_{i})]
\eea 

The $\Phi^{j}$ s can be a set of GUT singlet supermultiplets 
$\Phi^{S}$, which are part of the hidden 
sector, or a set of non-singlet ones $\Phi^N$, fields associated 
with the spontaneous breakdown of the GUT  group to 
$SU(3)\times SU(2)\times U(1)$ . The non-trivial gauge kinetic 
function $f_{\a  \b}(\Phi^{j}) $ can be expanded in 
terms of the non-singlet components of the chiral superfields in the 
following way 

\bea
f_{\a \b}(\Phi^{j})= f_{0}(\Phi^{S})\delta_{\a
  \b}+\sum_{N}\xi_{N}(\Phi^s)
\frac{{\Phi^{N}}_{\a \b}}{M}+ {\mathcal{O}}(\frac{\Phi^N}{M})^2
\label {gkf}
\eea
\noindent
where $f_0$ and $\xi^N$ are functions of chiral singlet superfields,
essentially determining the strength of the interaction and
$M$ is the reduced Planck mass$=M_{Pl}/\sqrt{8\pi}$.

In Equation (\ref {gkf}), the contribution to the gauge kinetic function
from  $\Phi^{N}$ has to come through symmetric products of the 
adjoint representation of the associated GUT group, since $f_{\a \b}$ on the 
left side of Equation (\ref {gkf}) has such transformation property for the sake of gauge 
invariance.

\noindent
For $SO(10)$, one can have contributions to $f_{\a \b}$ from all possible 
non-singlet irreducible representations to which $\Phi^{N}$ can belong :\\
\bea
(45\times 45)_{symm}=1+54+210+770
\eea
As an artifact of the expansion of the gauge kinetic
function $f_{\a \b}$ mentioned in Equation (\ref {gkf}), the gauge kinetic term 
(2nd term) in the Lagrangian (Equation (\ref {lag})) can be 
recast in the following form 

\bea
 Re f_{\a \b}(\phi)F_{\mu \nu}^{\a}F^{\beta \mu \nu}= \frac {\xi_{N}(\Phi^s)}
{M}Tr(F_{\mu \nu}\Phi^N F^{\mu \nu})
\eea
\noindent
where $F_{\mu \nu}$, under unbroken $SO(10)$, contains $U(1)_Y$, 
$SU(2)_L$ and $SU(3)_C$ gauge fields.

Next, the kinetic energy terms are restored to the canonical form
by rescaling the gauge superfields, by defining

\be 
{F^{\a}}_{\mu \nu} \rightarrow {\hat{F}^{\a}}_{\mu \nu}={\langle Re f_{\a
    \b} \rangle}^{\frac{1}{2}}{F^{\b}}_{\mu \nu} 
\label {res1}
\ee
and    
\be
\l^{\a}  \rightarrow {\hat{\l}}^{\a}={\langle Re f_{\a
    \b} \rangle}^{\frac{1}{2}}\l^{\b}
\label {res2}
\ee  
Simultaneously, the gauge couplings are also rescaled (as a result of Equation (\ref {gkf})):
\be
g_{\a}(M_{X}){\langle Re f_{\a \b} \rangle}^{\frac{1}{2}}\delta_{\a \b}= 
g_{c}(M_{X}) 
\label {res3}
\ee
where $g_{c}$ is the universal coupling constant at the GUT scale ($M_{X}$). This
shows clearly that the first consequence of a non-trivial
gauge kinetic function is non-universality of the gauge couplings $g_{\a}$ 
at the GUT scale \cite{Ellis,Drees,Hill,Shafi}.

Once SUSY is broken by non-zero $vev$'s of the $\tilde{F}$ components of 
hidden sector chiral superfields, the coefficient of 
the last term in Equation (\ref {lag}) is replaced by \cite{Ellis,Drees}
\be
{\langle { \tilde{F}_{\a \b}}^{i} \rangle}= {\mathcal{O}}(m_{\frac{3}{2}} M)
\ee
where $m_{\frac{3}{2}}= exp(-\frac{\langle G\rangle}{2})$ 
is the gravitino mass. Taking into account the rescaling of the gaugino fields 
(as stated earlier in Equation (\ref {res2})) in Equation (\ref {lag}), 
the gaugino mass matrix can be written down as \cite{Ellis,Pran} 
\be
M_{\a}(M_{X})\delta_{\a \b}=\sum_{i}\frac{{\langle F^{i}_{\acute{\a}
      \acute{\b}}\rangle}}{2} \frac{\langle \partial{f_{\a
    \b}(\phi^{*i})}/\partial{{\phi^{*i}}_{\acute{\a} \acute{\b}}} \rangle}
{\langle Re f_{\a \b} \rangle}
\ee
which demonstrates that the gaugino masses are non-universal at the GUT scale.

The underlying reason for this is the fact that  
{\large$\langle f_{\a \b} \rangle $} can 
be shown to acquire the form {\large $ f_{\a}\delta_{\a \b}$}, 
where the $ f_{\a}$ 's are purely group theoretic factors, as we will see. 
On the contrary, if symmetry breaking occurs via gauge
singlet fields only, one has {\large$f_{\a \b}=f_{0}\delta_{\a \b}$} 
from Equation (\ref {gkf}) and as a result, {\large $\langle f_{\a \b} \rangle=f_{0}$}. 
Thus both gaugino masses and the gauge couplings are unified at the GUT scale 
(as can be seen from Equations (8) and (10)). 

As mentioned earlier, we would like to calculate here, the $ f_{\a}$ 's 
for Higgses ($\Phi^N$) belonging to the representations {\bf 54} and {\bf 770} 
which break $SO(10)$ to the intermediate gauge group $ SU(4)_C \times SU(2)_L \times SU(2)_R $ with unbroken $D$-parity (usually denoted as $G_{422P}$) \footnote{Higgs, that breaks $G_{422P}$ to SM, 
doesn't contribute to gaugino masses.}. We associate the non-universal contributions to the gaugino mass ratios  with the group theoretic coefficients $ f_{\a}$ 's that arise here. It in turn, indicates that we consider the non-universality in the gaugino masses of $O(1)$. This, however, is not a generic situation in such models. This 
can be rather achieved under some special conditions like dynamical generation of SUSY-breaking scale from the electroweak scale, no soft-breaking terms for the GUT or Planck scale particles and with the simplified assumption 
$M_{GUT}=M_{Pl}$ which is also reflected in the RGE specifications.

The representations of $SO(10)$ \cite {Slansky},
decomposed into that of the Pati-Salam gauge group are
\bea 
SO(10) & \rightarrow & G_{422} = SU(4)_C \times SU(2)_L \times SU(2)_R \nonumber \\
 {\bf 45} & \rightarrow & ({\bf 15},{\bf 1},{\bf 1})
\oplus ({\bf 1},{\bf 3},{\bf 1}) \oplus ({\bf 1},{\bf 1},{\bf 3})
\oplus ({\bf 6},{\bf 2},{\bf 2}) \nonumber \\
{\bf 10} & \rightarrow & ({\bf 6},{\bf 1},{\bf 1}) \oplus
({\bf 1},{\bf 2},{\bf 2}) \\
{\bf 16} & \rightarrow & ({\bf 4},{\bf 2},{\bf 1}) \oplus ({\bf
\bar{4}},{\bf 1},{\bf 2}) \nonumber
\eea

Using the $SO(10)$ relation, ($10 \times 10$) = $1+ 45+54$,  
one can see that $vev$ of 54-dimensional Higgs (${\langle 54 \rangle}$) can be expressed as a 
(10 $\times$ 10) diagonal traceless matrix.
Thus the non-zero $vev$ of {\bf 54}-dimensional Higgs can be written as \cite{Shafi}
\bea
<54>=\frac{v_{54}}{2\sqrt{15}}~diag(3,3,3,3,,-2,-2,-2,-2,-2,-2).
\label{ve54}
\eea

Since $(45 \times 45)_{sym} = 1+54+210+770 $, 
one can write the non-zero $vev$ \cite{Joydeep} of {\bf 770}-dimensional Higgs
as ($45 \times 45$) diagonal matrix:
\bea
{\langle 770 \rangle}= \frac {v_{770}}{\sqrt{180}}~ diag (\underbrace 
{-4,......,-4}_{15},\underbrace {-10,...,
-10}_{3+3},\underbrace {5,......,5}_{24})
\eea

In the intermediate scale ($M_C$), $G_{422}$ is broken to SM gauge group.
Here $SU(4)_C$ is broken down to $SU(3)_C \times U(1)_{B-L}$
and at the same time, $SU(2)_R$ is broken to $U(1)_{T_3R}$.
It is noted that $SU(2)_R \times SU(4)_C$ is broken to $SU(3)_C \times U(1)_Y$
and hence the hypercharge is given as $Y  =  T_{3R} + \frac{1}{2}(B-L)$.
Below we note the branchings of $SU(4)_C$ representations:
\bea
SU(4)_C & \rightarrow & SU(3)_C \times U(1)_{B-L} \nonumber \\
{\bf 4} & \rightarrow & {\bf 3}_{1/3} \oplus {\bf 1}_{-1} \nonumber \\
{\bf 15} & \rightarrow & {\bf 8}_0 \oplus {\bf 3}_{4/3}
\oplus {\bf \bar{3}}_{-4/3} \oplus {\bf 1}_0  \\
{\bf 10} & \rightarrow & {\bf 6}_{2/3}
\oplus {\bf 3}_{-2/3} \oplus {\bf 1}_{-2} \nonumber \\
{\bf 6} & \rightarrow & {\bf 3}_{-2/3} \oplus {\bf \bar{3}}_{2/3} \nonumber
\eea 

Combining these together, we achieve the branchings of $SO(10)$
representations in terms of the SM gauge group. 
\bea
& SO(10) & : SU(3)_C \times SU(2)_L \times U(1)_Y  \nonumber\\
& {\bf 45} & : \nonumber \\
& ({\bf 15},{\bf 1},{\bf 1}) & \rightarrow ({\bf 8},{\bf 1})_0
\oplus ({\bf 3},{\bf 1})_{2/3} \oplus ({\bf \bar{3}},{\bf
1})_{-2/3} \oplus ({\bf
1},{\bf 1})_0 \nonumber \\
 & ({\bf 1},{\bf 3},{\bf 1}) & \rightarrow ({\bf 1},{\bf 3})_0 \nonumber \\
 & ({\bf 1},{\bf 1},{\bf 3}) & \rightarrow ({\bf 1},{\bf 1})_1
 \oplus ({\bf 1},{\bf 1})_0 \oplus ({\bf 1},{\bf 1})_{-1} \nonumber \\
 & ({\bf 6},{\bf 2},{\bf 2}) & \rightarrow ({\bf 3},{\bf 2})_{1/6}
 \oplus ({\bf 3},{\bf 2})_{-5/6} \oplus
({\bf \bar{3}},{\bf 2})_{5/6} \oplus ({\bf \bar{3}},{\bf 2})_{1/6} \\
 & {\bf 10} & :  \nonumber \\
 & ({\bf 6},{\bf 1},{\bf 1}) & \rightarrow ({\bf 3},{\bf 1})_{-1/3}
 \oplus ({\bf \bar{3}},{\bf 1})_{1/3} \nonumber \\
 & ({\bf 1},{\bf 2},{\bf 2}) & \rightarrow ({\bf 1},{\bf 2})_{1/2}
 \oplus ({\bf 1},{\bf 2})_{-1/2}  \\
 & {\bf 16} & :  \nonumber \\
 & ({\bf 4},{\bf 2},{\bf 1}) & \rightarrow ({\bf 3},{\bf 2})_{1/6}
 \oplus ({\bf 1},{\bf 2})_{-1/2} \nonumber \\
 & ({\bf \bar{4}},{\bf 1},{\bf 2}) & \rightarrow
 ({\bf \bar{3}},{\bf 1})_{1/3} \oplus
({\bf \bar{3}},{\bf 1})_{-2/3} \oplus ({\bf 1},{\bf 1})_1 \oplus
({\bf 1},{\bf 1})_0  
\eea

We have $U(1)_{T_{3R}}$ and $U(1)_{B-L}$ from $SU(2)_R$ and $SU(4)_C$ respectively.
Thus the weak hyper-charge generator ($T_{Y}$) can be expressed as a linear combination
of the generators of $SU(2)_R$ ($T_{3R}$) and $SU(4)_C$ ($T_{B-L}$) 
sharing the same quantum numbers.
In 10-dimensional representation $T_{3R}$, $T_{B-L}$ and $T_Y$ are written as:
\bea
T_{3R}=diag(0,0,0,0,0,0,\frac{1}{2},-\frac{1}{2},\frac{1}{2},-\frac{1}{2})
\eea
\bea
T_{B-L}=\sqrt{\frac{3}{2}} diag(-\frac{1}{3},-\frac{1}{3},-\frac{1}{3},\frac{1}{3},\frac{1}{3},\frac{1}{3},0,0,0,0)
\eea
\bea
T_Y=\sqrt{\frac{3}{5}} diag(-\frac{1}{3},-\frac{1}{3},-\frac{1}{3},\frac{1}{3},\frac{1}{3},\frac{1}{3},
\frac{1}{2},-\frac{1}{2},\frac{1}{2},-\frac{1}{2})
\eea

Using these explicit forms of the generators we find the following relation
\bea
T_{Y}=\sqrt{\frac{3}{5}} T_{3R}+\sqrt{\frac{2}{5}} T_{B-L}
\eea
and this leads to the following mass relation,
\bea
M_1 =\frac {3}{5}~M_{2R} + \frac {2}{5}~M_{4C}  
\label {mr1}
\eea
\begin{itemize}
\item For {\bf 54}-dimensional Higgs:
\end{itemize}
Using {\bf 54}-dimensional Higgs we have \cite{Shafi}, for $D$-parity even scenario, 
$M_{4C}=1$ and $M_{2R}=M_{2L}=-\frac {3}{2}$.
We have identified that $M_3=M_{4C}$ and $M_2=M_{2R}$. Hence, using the above 
mass relation we obtain $M_1= -\frac{1}{2}$.
Therefore the gaugino mass ratio is given as:
\bea
M_1:M_2:M_3= (-\frac{1}{2}):(-\frac{3}{2}):1
\eea

We have already mentioned that the $vev$ \cite{Joydeep} of {\bf 770}-dimensional Higgs
can be expressed as a ($45 \times 45$) diagonal matrix.
So to calculate the gaugino masses, using {\bf 770}-dimensional Higgs, we repeat 
our previous task in 45-dimensional representation. The explicit forms of 
$T_Y$, $T_{B-L}$ and  $T_{3R}$ are noted in the Appendix and find the same 
mass relation as in Equation (\ref {mr1})
\bea
M_1 =\frac {3}{5}~M_{2R} + \frac {2}{5}~M_{4C}
\label {mr2}
\eea
This mass relation among the $U(1)_Y$, $SU(2)_L$ and $SU(3)_C$ 
gauginos are independent of dimensions and representations.

\begin{itemize}
\item {\bf 770}-dimensional Higgs:
\end{itemize}
Using {\bf 770}-dimensional Higgs we find \cite{Joydeep} for $D$-parity even case
$M_{4C}=2$ and $M_{2R}=M_{2L}=5$.
Hence, using the above mass relation we obtain $M_1= 3.8$.
Therefore the gaugino mass ratio is given as:
\bea
M_1:M_2:M_3= 1.9:2.5:1
\eea

We tabulate the gaugino mass ratios, obtained above, in Table 1.
\vspace{0.2cm}
\begin{center}
\begin{tabular}{|c|c|}
\hline
\hline
 Representation & $M_{3}:M_{2}:M_{1}$ at $M_{GUT}$ \\
\hline
{ 1} & 1:1:1 \\
\hline
{\bf 54}: {$H \rightarrow SU(4) \times SU(2) \times SU(2)$} & 1:(-3/2):(-1/2)\\
\hline
{\bf 770}: {$H \rightarrow SU(4) \times SU(2) \times SU(2)$} & 1:(2.5):(1.9)\\
\hline
\hline
\end {tabular}\\
\noindent
\end {center}
\vspace {0.2cm}
\begin{center}
{Table 1: {\em High scale gaugino mass ratios for the
representations {\bf 54} and {\bf 770}.}}\\
\end{center}

Now we would like to address the question whether 
{\em the $D$-Parity Odd case is phenomenologically viable.}
We know, if we use 210-dimensional Higgs instead of {\bf 54} or {\bf 770}, 
the $D$-parity is broken by $vev$ of 210 dimensional Higgs when 
$SO(10) \longrightarrow SU(4)_C \times SU(2)_L \times SU(2)_R$. This has a nice feature:
$M_{2L}=-M_{2R}$ and $M_{4C}=0$, which implies that at high scale $M_3$
will be zero. We have analyzed the running of such high scale parameters numerically 
and found that these cases are not phenomenologically viable.
With $M_3=0$ at high scale, we will, always, hit Landau pole i.e. we will be in non-perturbative 
regime after running down to the low scale. 
{\em This is not only the specialty of 210-dimensional Higgs, but also true for all 
the $D$-parity breaking scenarios}. Hence we conclude that the $D$-parity
non-conserving scenario doesn't fit into the non-universal gaugino mass framework, 
in particular, when we are bothered about a MSSM spectrum valid at the EWSB scale.

\subsection{Implication of the Intermediate Scale}

We have an underlying assumption that the breaking of $SO(10)$ GUT group to the intermediate
gauge group and that to the SM takes place at the GUT scale itself, which is of course a 
simplification. But more interesting question to ask is {\em how things will 
change if the intermediate scale is different from the GUT scale 
(which is usually the most realistic one)?} Although we do not address this 
question in this analysis, we prefer to mention the crucial consequences of 
choosing an intermediate scale distinctly different from GUT scale:
\bei
\item In this case, the choice of the non-singlet Higgses will be restricted.
Now, only those Higgses will contribute which have a singlet direction
under the intermediate gauge group. Hence, the Higgses that break $SO(10)$
directly to the SM at the GUT scale are disallowed.
The choices of the non-singlet Higgses, in our analysis, are compatible
even if the intermediate scale is different from the GUT scale.
\item The mass relation in Equation (\ref{mr1}) is indeed independent  
of the intermediate scale as it is an outcome of purely group theoretical
analysis. But the gaugino mass ratios will change depending on the choice of 
the intermediate scale due to the running of the gaugino masses from the GUT scale.
\eei
      
\section {Low energy spectra, Consistency and Benchmark Points}

Before we discuss in details the low-energy spectra for the non-universal 
inputs, we would like to mention a few points regarding the evolution of 
these gaugino mass ratios with different RGE specifications. 
As we know, in the one-loop RGE, the gaugino mass parameters do not 
involve the scalar masses \cite{Rammond}, the ratios obtained at the low scale
are independent of the high scale scalar mass input $m_0$. In addition, if 
we also assume no radiative corrections (R.C)\footnote{The radiative 
corrections to the gauginos have been incorporated in calculating the 
physical masses after the RGE using the reference \cite{Pierce}.} to the gaugino 
masses, the ratios at the EWSB scale are also independent of the 
choice of the gaugino mass parameters at the high scale. Instead, if one uses 
the two-loop RGE (scalars contributing to the gauginos), the values of the 
gaugino masses at the EWSB scale  tend to decrease compared to the values 
obtained with one-loop RGE. Now, independently, the inclusion of R.C to the 
gaugino masses, makes the $M_3$ lower, but the values of $M_1$ and $M_2$ become 
higher compared to the case of one-loop results with no R.C. When, one uses 
both the two-loop RGE and R.C to the gauginos, it is a competition between 
these two effects. In short, the gaugino mass ratios at the EWSB scale crucially 
depend on the choice of the RGE specifications. However, the dependence on the 
high scale mass parameters $m_0$ and/or $M_3$ is very feeble\footnote {In 
particular, with change in $M_3$ from 300-1000 GeV, with $m_0$=1000 GeV, the 
change in the ratios is within 10$\%$, where as the ratios remain almost the same
with change in $m_0$.}.

We present in Table 2, the gaugino mass 
ratios at the EWSB scale for two different RGE conditions: 
\begin{itemize}
\item One-loop RGE with no R.C to the gaugino masses
\item Two-loop RGE with R.C to the gaugino masses
\footnote{The case with two-loop RGE + R.C to the gaugino masses have been 
obtained with $m_0=M_3$= 500 GeV.}
\end{itemize}

\begin{center}
\begin{tabular}{|l|c|c|}

\hline
\hline
 Representation & $M_{3}:M_{2}:M_{1}$ &  $M_{3}:M_{2}:M_{1}$\\
   & one-loop with No R.C & two-loop with R.C\\
\hline
{ 1 ({\bf mSUGRA})} & 1:0.27:0.13 & 1:0.35:0.19 \\
\hline
{\bf 54}: {$H \rightarrow SU(4) \times SU(2) \times SU(2)$} & 
1:(-0.40):(-0.06) & 1:(-0.55):(-0.10) \\
\hline
{\bf 770}: {$H \rightarrow SU(4) \times SU(2) \times SU(2)$} & 1:0.67:0.24 
& 1:0.91:0.37\\
\hline
\hline
\end {tabular}\\
\noindent
\end {center}
\vspace{0.2cm}
\begin{center}
{Table 2: {\em Low scale (EWSB) gaugino mass ratios 
for representations {\bf 54} and {\bf 770}.}}\\
\end{center}

The numerical results have been obtained using the spectrum generator 
{\tt SuSpect v2.3} \cite{SuSpect} with the {\bf pMSSM} option.
For the rest of our analysis we adhere to the second type of RGE specifications, 
{\em two-loop RGE + R.C to the gauginos} as mentioned earlier.
The other broad specifications used for the scanning are listed below.

\begin{itemize}
\item Full one-loop and the dominant two-loop 
corrections to the Higgs masses are incorporated.
\item Gauge coupling constant unification at the high scale have been ensured
and the corresponding scale has been chosen as the 'high scale' or 'GUT-scale' to
start the running by RGE. All the non-universal inputs are provided at this scale
using the {\bf pMSSM} option. This is an artifact of choosing the intermediate scale 
set at the GUT scale itself.
\item Electroweak symmetry breaking at the `default scale'  
$\sqrt{m_{\tilde{t_{L}}}m_{\tilde{t_{R}}}}$ has been set.
\item We have used the strong coupling ${\alpha_3 (M_{Z})}^{\overline{MS}}= 0.1172$ 
for this calculation which is again the default option in {\tt SuSpect}.
\item Throughout the analysis we have assumed the top quark mass to be 171.4 GeV.
\item All the scalar masses have been set to a universal value of $m_0$ and 
radiative electroweak symmetry breaking has been taken into account by setting 
high scale Higgs mass parameter ${M_{H_u}}^2={M_{H_d}}^2={m_0}^2$ and specifying 
$sgn(\mu)$, which has been taken to be positive throughout the analysis.
\item All the trilinear couplings have been set to zero.
\item Tachyonic modes for sfermions and other inconsistencies in RGE, like Landau 
pole have been taken into account.
\item As we work in a $R$-parity conserving scenario, stau-LSP regions have been
identified as disfavoured.
\item Consistency with low-energy FCNC constraints such as those from 
$b\rightarrow s\gamma$ has been noted for each combination of the parameter
space. We have used a 3$\sigma$ level constraint from $b \rightarrow s \gamma $ 
with the following limits \cite{bsg-recent}. 
\begin{equation}
2.77 \times 10^{-4} < Br (b \rightarrow s \gamma) < 4.33 \times 10^{-4}.
\label{bsgammalimits}
\end{equation}
However, we must point out that we have taken all those regions as allowed 
where the value of $b\rightarrow s\gamma$ is lower or within the constraint. 
\item Regions allowed by all these constraints have been studied for the  
relic density constraint of the cold dark matter (CDM) candidate 
(lightest neutralino in our case)  and referred to the  
the WMAP data \cite{WMAPdata} within 3$\sigma$ limit
\begin{equation}
0.091 < \Omega_{CDM}h^2 < 0.128.
\label{relicdensity}
\end{equation}
where $\Omega_{CDM}h^2$ is the dark matter relic density in units of the 
critical density and $h=0.71\pm0.026$ is the Hubble constant in units of
$100 \ \rm Km \ \rm s^{-1}\ \rm Mpc^{-1}$.
\end{itemize}
We have used the code {\tt microOMEGA v2.0.7} \cite{micromegas} for computing 
the relic density.

With these inputs, we scan the parameter space for a wide range of values 
of $m_0$ and $M_3$ \footnote {Choice of $M_3$ automatically determines the 
values of $M_1$ and $M_2$ for a choice of non-universality.} for the 
non-universal gaugino mass ratios advocated above.

The ratios obtained for {\bf 54} at the high scale (see Table 1) are 
actually the same as the one for the representation {\bf 24} in case of 
$SU(5)$ (see \cite{Ellis,Drees,Pran,Subho}). This observation differs from the 
earlier result available in \cite{Chamoun}. The low-energy spectrum and 
its consistency for the case of {\bf 24} have been well-studied 
\cite{24}. Without the inclusion of the intermediate breaking scale 
in case of $SO(10)$, the case of {\bf 54} is difficult to distinguish from the 
one in $SU(5)$. Anyway we do not address any such situations here and hence, 
refrain from illustrating the case of {\bf 54}.

In Figure 1, we depict the results of the scan in the $M_3-m_0$ parameter 
space for the representation {\bf 770}, i.e. breaking through $G_{422}$.
Along the x-axis, high scale $M_3$ is varied from 100-2000 GeV and 
along the y-axis, high scale universal scalar mass $m_0$ is varied in the same range.
Our limit of the scan is motivated by the fact that we cover the low scale parameters 
well beyond the reach of the LHC. The figure on the left hand side is 
for $\tan\beta$= 5 and on the right hand side is for $\tan\beta$= 40.

For $\tan\beta$= 5, full parameter space is allowed by REWSB, 
$b \rightarrow s\gamma$ and other RGE constraints. The black region
at the bottom (for $M_3$= 400-1100 GeV and for very small values of $m_0$) is
disfavoured by the stau-LSP constraint. Hence, there is a large region 
of the parameter space (shown in red) which satisfies all the constraints 
and it is definitely within the reach of the LHC. We study the dark-matter 
constraints in this allowed region of parameter space and our conclusion 
is as follows: 
\begin{itemize}
\item For $M_3$= 200 GeV, the allowed 
range of $m_0$ spans around 200 GeV
\item For $M_3$= 400 GeV, the allowed
region is extremely narrow (because of the stau-LSP constraint) and is around  
$m_0$= 140 GeV
\item For $M_3$= 600 GeV, $m_0$= 300 GeV is allowed 
\item For $M_3$= 800 GeV, the value of $m_0$ goes as high as 1100 GeV
\end{itemize}
We choose three benchmark points (BP1, BP2 and BP3, see Table 2 and 3) from here 
and study the collider signature.

The figure on the right hand side of Figure 1 is with $\tan\beta$= 40 and is 
quite different from the one with $\tan\beta$= 5. In this figure, the red region 
is allowed by REWSB while the area in blue at small $m_0$ is disfavoured by the 
stau-LSP constraint. Hence, here also, there exists a large region of parameter 
space, sandwiched between these two, allowed by all the constraints for
the study of dark-matter and collider search. Excepting for a very narrow region 
at the left bottom corner spanning 100-200 GeV of $M_3$ or $m_0$ value (in green), 
the whole region is under the $b \rightarrow s\gamma$ upper limit. The dark 
matter study in this case, yields something special. We find almost all the regions 
to be lying below the lower bound of the WMAP data\footnote{This is not strictly 
disfavoured as some other scenarios beyond the SM can co-exist and contribute.}.
It is worthy to mention here that non-universal gaugino mass scenarios with $M_2>M_3$ at high-scale
yield smaller value of $\mu$ generated from REWSB for a fixed $M_1$ at TeV scale when compared with mSUGRA. 
This makes the neutralino-LSP more Higgsino like, increases the annihilation and consequently gives a larger region of the parameter space compatible with dark matter constraints \cite{DM2}.
We choose a couple of benchmark points (BP4 and BP5, see Table 2 and 3) from here 
for the collider study.      

\begin{figure}[htbp]
\begin{center}
\centerline{\psfig{file=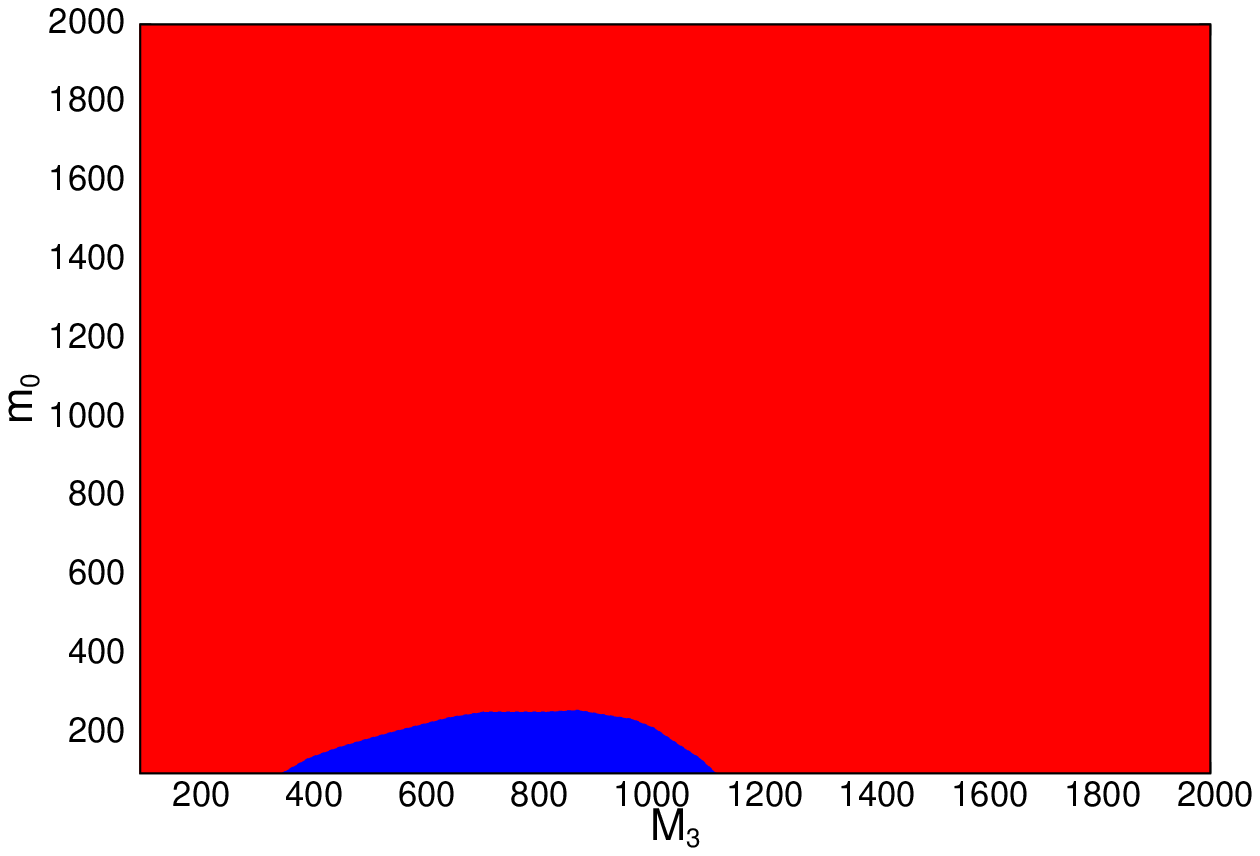,width=7.5 cm,height=6.5cm,angle=0}
\hskip 20pt \psfig{file=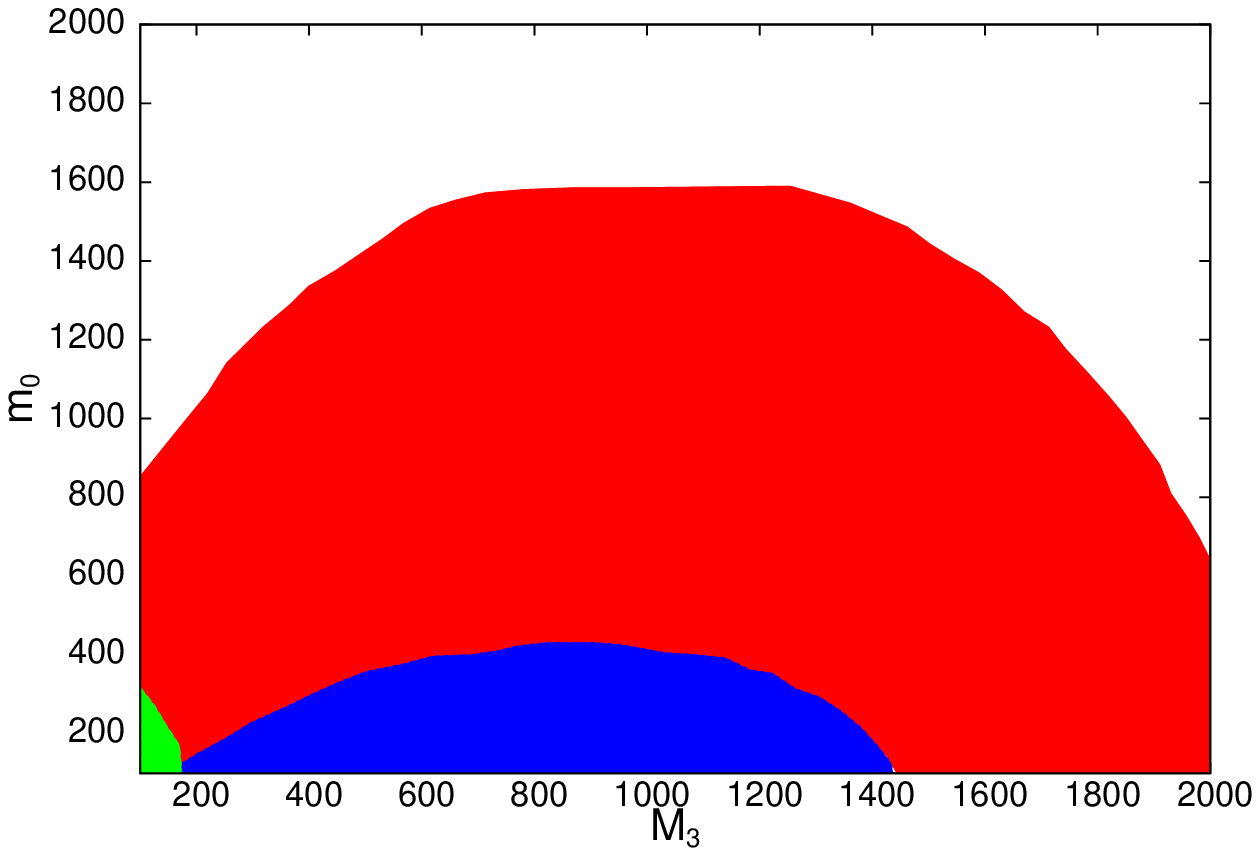,width=7.5 cm,height=6.5cm,angle=0}}
\caption{{\em Parameter space scan for the representation 
{\bf 770} (breaking through $G_{422}$) at high scale $M_3-m_0$ plane (in GeV), showing 
the regions allowed by various constraints. {\tt Figure on the left hand side}: 
{$\tan \beta$= 5. The red region is allowed and the blue region at the 
bottom is disfavoured by stau-LSP.} {\tt Figure on the right hand side}: 
{ $\tan \beta$= 40. The region in red is allowed by all constraints. Region 
in white is disfavoured by REWSB, the low-$m_0$ region, in blue, is disfavoured by 
stau-LSP and the left corner of the graph in green with small $M_3$ and $m_0$ is 
disfavoured by $b \rightarrow s\gamma$.}}} 
\end{center}
\end{figure}

Figure 2, shows similar parameter space scan for the case of mSUGRA, for 
$\tan\beta$= 5 (left) and $\tan\beta$= 40 (right). These are presented to show 
the difference in the low-energy parameter space consistency patterns for 
different high scale gaugino mass inputs. These cases have been studied well 
and need not require much illustration. Our scan seems to match with the earlier available 
results \cite {nonugmpheno2} and show the robustness of our analysis. However, it is worth 
mentioning that the red region, excepting for the blue region at small 
$m_0$ (disfavoured by stau-LSP), is allowed for $\tan\beta$= 5. For $\tan \beta$=40, 
a small region at the upper left corner with high values of $m_0$ (1200-2000 GeV) and 
small $M_3$ (100-400 GeV) (in white) is disfavoured by REWSB. The blue region at the 
bottom is disfavoured for stau-LSP.

\begin{figure}[htbp]
\begin{center}
\centerline{\psfig{file=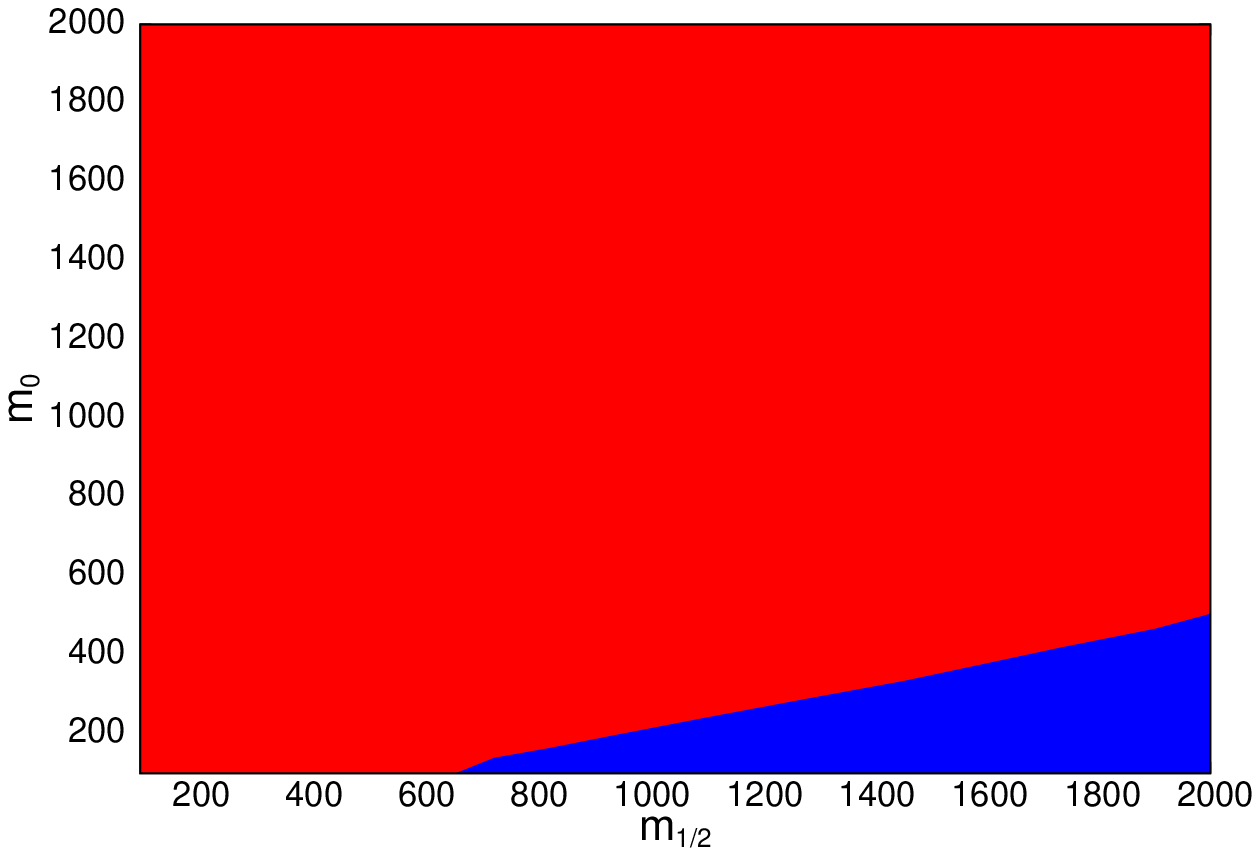,width=7.5 cm,height=6.5cm,angle=0}
\hskip 20pt \psfig{file=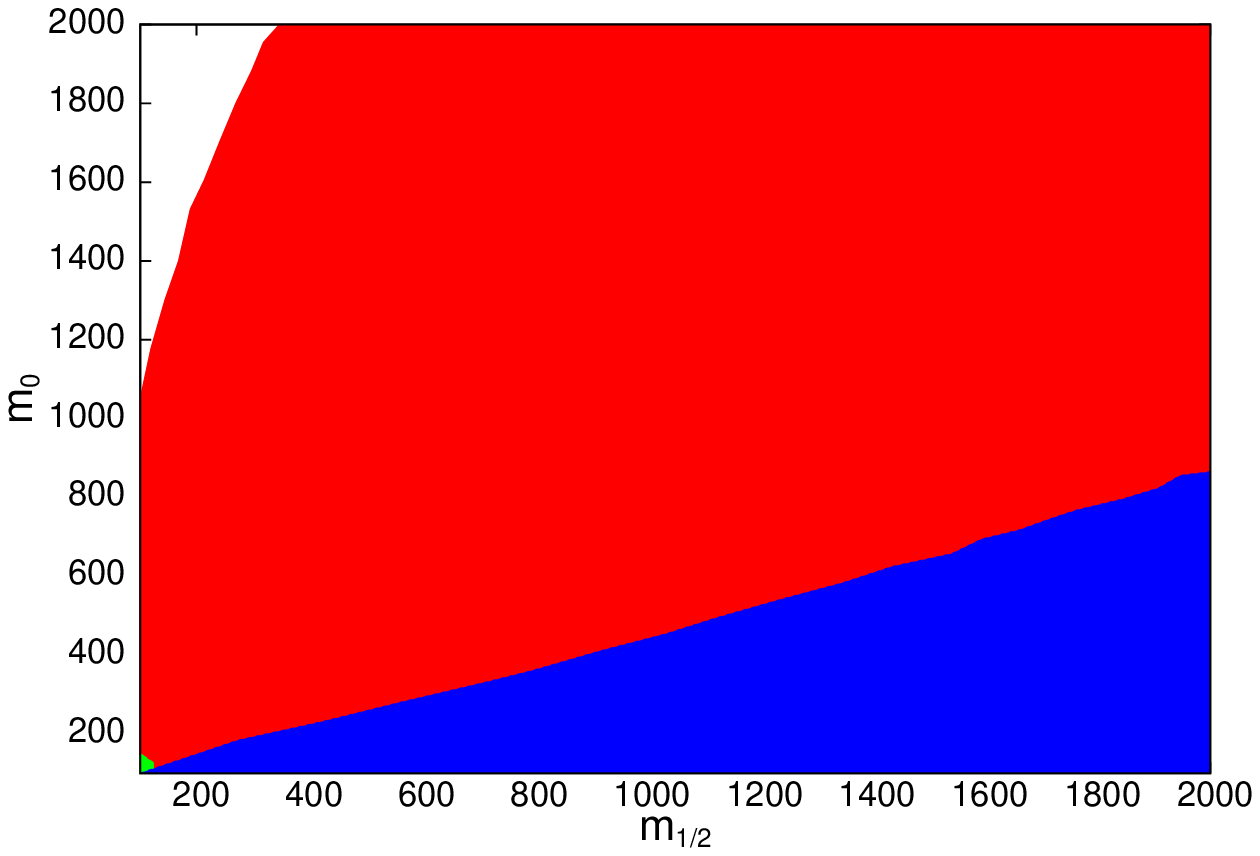,width=7.5 cm,height=6.5cm,angle=0}}
\caption{{\em Parameter space scan for {\bf mSUGRA} at high scale $m_{1/2}-m_0$ plane 
(in GeV), showing the regions allowed by various constraints.
{\tt Figure on the left side}: { $\tan \beta$= 5. The red region is allowed. 
The blue region at the bottom with low $m_0$ is disfavoured by stau-LSP.} 
{\tt Figure on the right side}: {$\tan \beta$= 40. The red region is allowed. 
The region in white at the left corner with high values of $m_0$ (1200-2000 GeV) 
is disallowed by REWSB and the bottom of the figure in blue with low values of $m_0$ 
(100-600 GeV), is disfavoured by stau-LSP.}}} 
\end{center}
\end{figure}

The benchmark points (BPs) chosen from Figure 1, to study the collider signature in context 
of the LHC, are presented in Table 3 and 4. In Table 3, we mention the high scale input 
parameters, while the low-energy spectra corresponding to these points have been 
mentioned in Table 4. The points have been chosen for two different values of $\tan \beta$, 
5 (BP1, BP2, BP3) and 40 (BP4, BP5) and have gluino masses around 500 GeV and 1000 GeV.
These points for $\tan \beta$= 5 satisfy the WMAP data for the cold dark matter relic 
density search, while the points for $\tan \beta$= 40 are all below the lower limit quoted 
by WMAP. The corresponding values of  $\Omega_{CDM}h^2$ have also been mentioned in Table 3. 
These points also obey the LEP bounds \cite{LEP}. The model under scrutiny has been referred 
as 770-422 in Table 3 and will be referred so in the following text.

\noindent

\begin{center}

\begin{tabular}{|c|c|c|c|c|c|}

\hline
\hline
 Benchmark Points & Model & $M_3$ & $m_{0}$ & $\tan \beta$ & $\Omega_{CDM}h^2$\\
\hline 
BP1 & 770-422 & 200 & 200 & 5 & 0.124 \\
\hline
BP2 & 770-422 & 400 & 140 & 5 & 0.125 \\
\hline
BP3 & 770-422 & 600 & 300 & 5 & 0.126 \\
\hline
BP4 & 770-422 & 200 & 200 & 40 & 0.0002 \\
\hline
BP5 & 770-422 & 400 & 700 & 40 & 0.0162 \\
\hline
\hline
\end {tabular}\\
\end {center}
\vspace{0.2cm}
\begin{center}
{Table 3: {\em Benchmark Points (BP): Models, High scale input parameters (in GeV), 
$\tan \beta$ and $\Omega_{CDM}h^2$.}}\\
\end{center}

\noindent

\begin{center}

\begin{tabular}{|l|c|c|c|c|c|c|c|}

\hline
\hline
 Benchmark Points & $m_{\tilde{g}}$ & $m_{\tilde {q}_{1,2}}$ &  $m_{\tilde {l}_{1,2}}$
 &$m_{\tilde {\chi}^{0}_{1}}$ & $m_{\tilde {\chi}^{\pm}_{1}}$ & $\mu$ \\
 & & $m_{\tilde {t}_{1}}$ & $m_{\tilde {\tau}_{1}}$& &$m_{\tilde {\chi}^{0}_{2}}$& \\

\hline

 BP1 & 499.95 & 524 & 319.5 & 138.43 & 203.5 & 221.4 \\
 & & 305.26 & 246.79 & & 219.56 & \\
\hline

 BP2 & 938.8 & 928 & 495 & 303.53 & 375.96 & 381.18 \\
 & & 551.83 & 313.93 & & 386.92 & \\
\hline
 BP3 & 1368.48 & 1366 & 770.5 & 464.3 & 515.65 & 515.27 \\
 & & 815.14 & 513.84 & & 522.39 & \\
\hline

 BP4 & 499.5 & 524.5 & 320 & 132.23 & 170.45 & 178.36 \\
 & & 315 & 172.25 & & 187.51 & \\
\hline
 BP5 & 966.58 & 1145 & 857 & 246.73 & 262.11 & 261.83 \\
 & & 699.06 & 597.87 & & 269.58 & \\
\hline
\hline 
\end {tabular}\\
\noindent
\end {center}
\vspace{0.2cm}
\begin{center}
{Table 4 : {\em Low-energy spectra for the chosen benchmark points (BP) (in GeV).}}\\
\end{center}

In Table 4, we note the gluino mass ($m_{\tilde{g}}$), average
 of the first two generation squark masses ($m_{\tilde {q}_{1,2}}$), average
 of the first two generation slepton masses ($m_{\tilde {l}_{1,2}}$), lighter
stau mass ($m_{\tilde {\tau}_{1}}$), lighter stop ($m_{\tilde {t}_{1}}$), 
lightest neutralino ($m_{\tilde {\chi}^{0}_{1}}$), lighter chargino 
($m_{\tilde {\chi}^{\pm}_{1}}$), 2nd lightest neutralino 
($m_{\tilde {\chi}^{0}_{2}}$) as well as the value of $\mu$, generated by REWSB at the BPs.
It can also be noted that BP1, BP2, BP3 have bino dominated $\tilde {\chi}^{0}_{1}$, 
while it is mixed and higgsino dominated in case of BP4 and BP5 respectively. 
The $\tilde {\chi}^{0}_{2}$ is mixed in case of BP1 and higgsino dominated in all 
the other cases. The lighter chargino is mostly higgsino dominated and degenerate with 
$\tilde {\chi}^{0}_{2}$, excepting for the case of BP1. We would like to mention that, 
the composition (gaugino dominated, higgsino dominated or mixed) and the mass difference of 
the neutralinos and charginos get altered for different high scale gaugino 
non-universality. Given the similar values of squark and gluino masses in 
different non-universal schemes (actually similar choices of $M_3$ and $m_0$ at the high scale), 
the electroweak gauginos become instrumental for a possible distinction between
different GUT-breaking schemes, which might also get reflected at the collider signature 
in a favourable region of parameter space.

To see such distinction in collider signature, we choose two points from mSUGRA scenario 
with suitable values of $M_{1/2}$ such that the low scale gluino masses are around 500 GeV 
and 1000 GeV (similar to the benchmark points (BP) selected above) in a region of 
parameter space that satisfies the WMAP data for cold dark matter constraint. 
It should be noted that once these points respect the CDM constraint, the value of 
$m_0$ automatically get restricted for a particular choice of $M_{1/2}$ for a specific 
$\tan \beta$. Similar is the situation here where we have taken $\tan \beta$ = 5 for 
illustration. The chosen points are named as MSG1 and MSG2. The high scale parameters 
along with the $\Omega_{CDM}h^2$ at these points are mentioned in Table 5, while the 
corresponding low scale spectra are noted in Table 6. 

\noindent

\begin{center}

\begin{tabular}{|c|c|c|c|c|c|}

\hline
\hline
 Points & Model & $m_{1/2}$ & $m_{0}$ & $\tan \beta$ & $\Omega_{CDM}h^2$\\
\hline 
 MSG1 & mSUGRA & 480 & 100 & 5 & 0.111 \\
\hline
 MSG2 & mSUGRA & 200 & 70  & 5 & 0.128 \\
\hline
\hline
\end {tabular}\\
\end {center}
\vspace{0.1cm}
\begin{center}
{Table 5: {\em mSUGRA points (MSG): Models, High scale input parameters (in GeV), 
$\tan \beta$ and $\Omega_{CDM}h^2$}. }\\
\end{center}

\vspace{0.1cm}
\begin{center}

\begin{tabular}{|l|c|c|c|c|c|c|c|}
\hline
\hline
 Points & $m_{\tilde{g}}$ & $m_{\tilde {q}_{1,2}}$ &  $m_{\tilde {l}_{1,2}}$
 &$m_{\tilde {\chi}^{0}_{1}}$ & $m_{\tilde {\chi}^{\pm}_{1}}$ & $\mu$ \\
 & & $m_{\tilde {t}_{1}}$ & $m_{\tilde {\tau}_{1}}$& &$m_{\tilde {\chi}^{0}_{2}}$& \\
\hline
 MSG1 & 1104.1& 992 & 272 & 195.66 & 367.1 & 622.1  \\
 & & 768.6 & 205.1 & & 367.4 & \\
\hline
 MSG2 & 493.25 & 452 & 129 & 73.2  & 131.4 & 281.4 \\
 & & 323.6 & 101.3 & & 133.6 & \\
\hline
\hline 
\end {tabular}\\
\noindent
\end {center}
\vspace{0.1cm}
\begin{center}
{Table 6 : {\em Low-energy spectra for the chosen mSUGRA points (MSG).}}
\end{center}

\section {Collider Simulation and Numerical Results}

We would like to discuss the collider signature now, 
of the benchmark points advocated in the preceeding section.

We first discuss the strategy for the simulation which includes the final state observables
and the cuts employed therein. In the next subsection we discuss the numerical results obtained
from this analysis. 
\subsection{Strategy for Simulation}
The spectrum generated by {\tt SuSpect} v2.3 as described in the earlier section, 
at the benchmark points are fed into the event generator 
{\tt Pythia} 6.4.16 \cite{PYTHIA} by {\tt SLHA} interface \cite{sLHA} 
for the simulation of $pp$ collision with centre of mass energy 14 TeV.

We have used {\tt CTEQ5L} \cite{CTEQ} parton distribution functions, 
the QCD renormalization and factorization scales being
both set at the subprocess centre-of-mass energy $\sqrt{\hat{s}}$.
All possible SUSY processes and decay chains consistent 
with conserved $R$-parity have been kept open. We have kept 
initial and final state radiations on. The effect of multiple 
interactions has been neglected. However, we take 
hadronization into account using the fragmentation functions 
inbuilt in {\tt Pythia}.

The final states studied here are :

\begin{itemize}
\item Opposite sign dilepton ($OSD$) :
$(\ell^{\pm}\ell^{\mp})+ (\geq 2)~ jets~ + {E_{T}}\!\!\!\!/$ 
  
\item Same sign dilepton ($SSD$) : 
$(\ell^{\pm}\ell^{\pm})+ (\geq 2)~jets~ + {E_{T}}\!\!\!\!/$

\item Trilepton $(3\ell+jets)$: 
$3\ell~ + (\geq 2) ~jets~ + {E_{T}}\!\!\!\!/$

\item Hadronically quiet trilepton  $(3\ell)$:
$3\ell~ + (0) ~jets~ + {E_{T}}\!\!\!\!/$   

\item Inclusive 4-lepton ($4\ell$): $4\ell + X + {E_{T}}\!\!\!\!/$    
\end{itemize}

\noindent
where $\ell$ stands for final state isolated electrons and or muons, 
${E_{T}}\!\!\!\!/$ depicts the missing energy, $X$ indicates any associated 
jet production.

We will discuss these objects in details, that constitute the final state 
observables. The nomenclature assigned to the final state events in parantheses
will be referred in the following text.

As defined in some earlier works \cite{Subho}, the absence of any jets with 
${E_{T}}^{jet} ~\geq ~100$ GeV qualifies the event 
as hadronically quiet. This avoids unnecessary removing of events along with 
jets originating from underlying events, pile up effects and ISR/FSR. 
The $4\ell$ events have been defined without putting an exclusive jet veto.

Before we mention the selection cuts, we would like to discuss the resolution 
effects of the detectors, specifically of the ECAL, HCAL and that of the muon 
chamber, which have been incorporated in our analysis. This is particularly 
important for reconstructing ${E_{T}}\!\!\!\!/$, which is a key variable for 
discovering physics beyond the standard model.  

All the charged particles with transverse momentum, $p_T~>$ 0.5 GeV\footnote 
{This is specifically for ATLAS, while for CMS, $p_T ~>$ 1 GeV is used.}   
that are produced in a collider, are detected due to strong B-field within a 
pseudorapidity range $|\eta|<5$, excepting for the muons where the range is 
$|\eta|<2.5$, due to the characteristics of the muon chamber.   
Experimentally, the main 'physics objects' that are reconstructed in a collider, 
are categorised as follows:
\begin{itemize}
\item Isolated leptons identified from electrons and muons
\item Hadronic Jets formed after identifying isolated leptons 
\item Unclustered Energy made of calorimeter clusters with $p_T~>$ 0.5 GeV 
(ATLAS) and $|\eta|<5$, not associated to any of the above types of 
high-$E_T$ objects (jets or isolated leptons).  
\end{itemize}  
Below we discuss the 'physics objects' described above in details.
\bei
\item {\em Isolated leptons} ($iso~\ell$):
\eei
Isolated leptons are identified as electrons and muons with $p_T>$ 10 GeV
and $|\eta|<$2.5. An isolated lepton should have lepton-lepton separation
${\bigtriangleup R}_{\ell\ell}~ \geq $0.2, lepton-jet separation 
(jets with $E_T >$ 20 GeV) ${\bigtriangleup R}_{{\ell}j}~ \geq 0.4$, 
the energy deposit $\sum {E_{T}}$ due to low-$E_T$ hadron activity around a 
lepton within $\bigtriangleup R~ \leq 0.2$ of the lepton axis should be 
$\leq$ 10 GeV, where  $\bigtriangleup R = \sqrt {{\bigtriangleup \eta}^2
+ {\bigtriangleup \phi}^2}$ is the separation in pseudo rapidity and 
azimuthal angle plane. The smearing functions of isolated electrons, photons and 
muons are described below.

\bei
\item {\em Jets} ($jet$):
\eei
Jets are formed with all the final state particles after
removing the isolated leptons from the list with {\tt PYCELL}, an inbuilt 
cluster routine in {\tt Pythia}. The detector is assumed to stretch within 
the pseudorapidity range $|\eta|$ from -5 to +5 and is segmented in 100 
pseudorapidity ($\eta$) bins and 64 azimuthal ($\phi$) bins. The minimum 
$E_T$ of each cell is considered as 0.5 GeV, while the minimum $E_T$ for
a cell to act as a jet initiator is taken as 2 GeV. All the partons within
$\bigtriangleup R$=0.4 from the jet initiator cell is considered for the jet 
formation and the minimum $\sum_{parton} {E_{T}}^{jet}$ for a collected cell 
to be considered as a jet is taken to be 20 GeV. We have used the smearing 
function and parameters for jets that are used in {\tt PYCELL} in {\tt Pythia}.

\bei
\item {\em Unclustered Objects} ($Unc.O$):
\eei
Now, as has been mentioned earlier, all the other final state particles,
which are not isolated leptons and separated from jets by 
$\bigtriangleup R \ge$0.4 are considered as unclustered objects. This clearly 
means all the particles (electron/photon/muon) with $0.5< E_T< 10$GeV and 
$|\eta|< 5$ (for muon-like track $|\eta|< 2.5$) and jets with $0.5< E_T< 20$GeV 
and $|\eta|< 5$, which are detected at the detector, are considered as 
unclustered energy and their resolution function have been considered separately
and mentioned below.

\begin{itemize}

\item {Electron/Photon Energy Resolution :}

\be
\sigma(E)/E=a/\sqrt{E}\oplus b\oplus c/E  \footnote{$\oplus$ indicates addition in quadrature}
\ee

Where,\\
$~~~~~~~~$ $a$ = 0.03 [GeV$^{1/2}$], $~~$   $b$ = 0.005 \& $~$   $c$ = 0.2 [GeV] 
$~~~$  for $|\eta|< 1.5$ \\     
$~~~~~~~~~~~~$= 0.055 $~~~~~~~~~~~~~~~~$ = 0.005  $~~~~~~~$ = 0.6 $~~~~~~~~~$  
for $1.5< |\eta|< 5$ \\

\item { Muon $P_T$ Resolution :}

\begin{eqnarray}
\sigma(P_T)/P_T&=&a  ~~~~~~~~~~~~~~~~~~~~~~~{\rm if} ~ P_T< 100 GeV\\
&=&a+b\log(P_T/\xi) ~~~~~ {\rm if}~ P_T> 100 GeV
\end{eqnarray}

Where,\\
$~~~~~~~~$ $a$= 0.008 $~~$\&    $b$= 0.037 $~~~~~$  for $|\eta|< 1.5$ \\     
$~~~~~~~~~~$= 0.02 $~~~~~~$    = 0.05  $~~~~~~~~~$ $1.5< |\eta|<2.5$

\item {Jet Energy Resolution :}

\be
\sigma(E_T)/E_T=a/\sqrt{E_T}
\ee

Where,\\
$~~~~~~~~$ a= 0.55 [GeV$^{1/2}$], default value used in {\tt PYCELL}.

\item {Unclustered Energy Resolution :}

\be
\sigma(E_T)=\a\sqrt{\Sigma_{i}E^{(Unc.O)i}_T}
\ee

Where, $\a\approx0.55$. One should keep in mind that the x and y component of 
$E^{Unc. O}_T$ need to be smeared independently with same smearing 
parameter.

\end{itemize}

All the smearing parameters that have been used are mostly in agreement
with the ATLAS detector specifications and also have been discussed in details
in \cite{Sanjoy}

Once we have identified the 'physics objects' as described above, we sum
vectorially the x and y components of the smeared momenta separately for isolated
leptons, jets and unclustered objects in each event to form visible transverse 
momentum $(p_T)_{vis}$, 
\bea
(p_T)_{vis}=\sqrt{(\sum p_x)^2+(\sum p_y)^2}
\eea
where, $\sum p_x =\sum (p_x)_{iso~\ell}+\sum (p_x)_{jet}+\sum (p_x)_{Unc.O}$
and similarly for $\sum p_y$.
We identify the negative of the $(p_T)_{vis}$ as missing energy $E_{T}\!\!\!\!/$:
\bea
E_{T}\!\!\!\!/ = -(p_T)_{vis}
\eea

Finally the selection cuts that are used in our analysis are as follows:

\begin{itemize}
\item  Missing transverse energy $E_{T}\!\!\!\!/$ $\geq ~100$ GeV. 

\item ${p_{T}}^\ell ~\ge ~20$ GeV for all isolated leptons.
 
\item ${E_{T}}^{jet} ~\geq ~100$ GeV and $|{\eta}_{jet}| ~\le ~2.5$  

\item For OSD, hadronically quiet trilepton ($3\ell$) and also for 
inclusive $4\ell$ events we have used, in addition, invariant mass cut on 
the same flavour opposite sign lepton pair as 
$|M_{Z}-M_{\ell_{+}\ell_{-}}| ~\geq 10$ GeV. 

\end {itemize}

We have checked the hard scattering cross-sections 
of various production processes with {\tt CalcHEP} \cite{CalcHEP}.
All the final states with jets at the parton level 
have been checked against the results available in \cite{Ash}. 
The calculation of hadronically quiet trilepton 
rates have been checked against \cite{hq3l}, in the appropriate limits.

We have generated dominant SM events in {\tt Pythia} for the same final 
states with same cuts. $t\bar t$ production gives the most serious 
backgrounds. We have multiplied the corresponding events in different channels 
by proper $K$-factor ($=2.23$) to obtain the usually noted next to leading order 
(NLO) and next to leading log resummed (NLL) cross-section of 
$t\bar t$ production at the LHC, 908 pb (without taking the PDF and scale uncertainty), 
for $m_t$ around 171 GeV \cite{ttbar}. The other sources of background include 
$WZ$ production, $ZZ$ production etc. The contribution of each of these 
processes to the various final states are mentioned in the Table 9.

\subsection{Numerical Results}
Figure 3 shows the {\em effective mass} distribution at the benchmark points
and the corresponding mSUGRA ones in $OSD$ events. Effective mass is defined as
\bea
Effective~mass=\sum (p_T)_{iso~\ell}+\sum (p_T)_{jets}+E_{T}\!\!\!\!/
\eea
Figure 3 has been organised following the model inputs. Top left figure shows
the distributions at BP1, BP2 and BP3 chosen from 770-422 with $\tan \beta$=5, 
whereas the top right one contains BP4 and BP5 chosen from the same scenario with 
$\tan \beta$=40 and the one from $t\bar t$ production, the dominant process for 
the background. The bottom one is for mSUGRA, containing MSG1 and MSG2 with $\tan \beta$=5.
The peak of the effective mass distribution corresponds to the {\tt threshold energy}
of the hard scattering process which is dominantly responsible for the final state 
under scrutiny. For $OSD$ events, processes responsible are mostly the $\tilde g \tilde g$, 
 $\tilde g \tilde q$ and $\tilde q \tilde q$ productions due to their $SU(3)$ interactions, 
provided they are accessible to the LHC center of mass energy. In such cases, the threshold energy
is around $2m_{\tilde g}$ or ($m_{\tilde g}+m_{\tilde q}$) or $2m_{\tilde q}$.
Now, in each of the BPs advocated here, $m_{\tilde g} =  m_{\tilde q}$ and the threshold
is approximately at $2m_{\tilde g}$. Our figures magnificently depict the correspondence 
with such threshold. For example, the peaks of BP1, BP4 and MSG2 are greater than 1000 GeV
(where the gluino and squark masses are around 500 GeV). The reason that these distributions peak 
at higher values than the threshold, can be attributed to the fact that the final state considered 
here, has a very large $E_{T}\!\!\!\!/$ cut, $(p_T)$ cut on associated jets and leptons.
While this indicates the robustness of our analysis, this also points to the deficiency 
in distinguishing these non-universal models from the mSUGRA one with similar gluino masses.
However, a possible way that could have been exploited is perhaps the effective mass 
distribution in $3\ell$ events. This is expected as the dominant production process
for this final state is $\tilde {\chi_2}^0$ and $\tilde {\chi_1}^{\pm}$ and these 
electroweak gauginos actually carry the information of different non-universal gaugino mass 
inputs at the GUT scale. However, this was not very successful in our case due to 
small event rates.

\begin{figure}[htbp]
\begin{center}
\centerline{\psfig{file=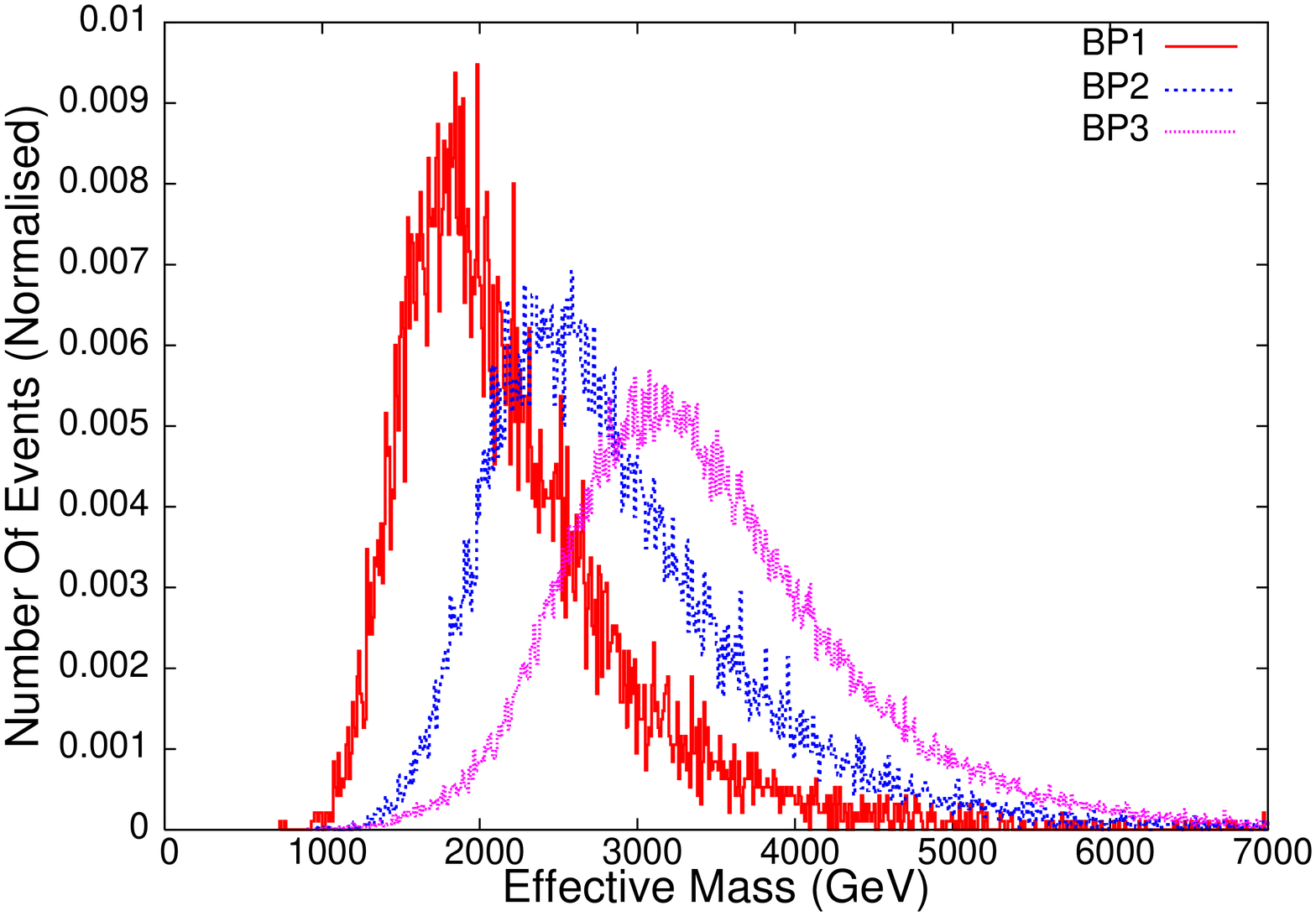,width=6.5 cm,height=7.5cm,angle=-90}
\hskip 20pt \psfig{file=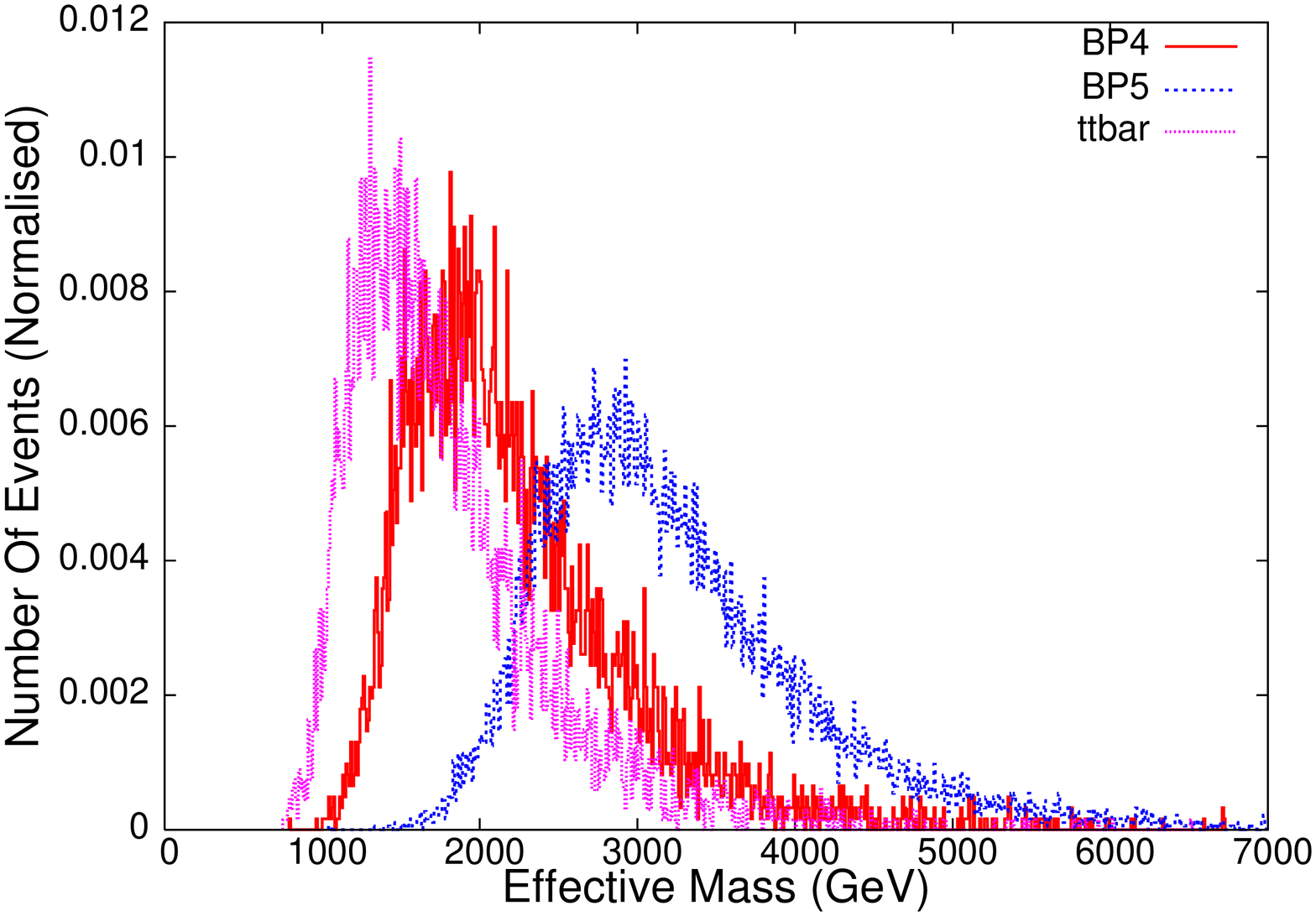,width=6.5 cm,height=7.5cm,angle=-90}}
\vskip 10pt
\centerline{\psfig{file=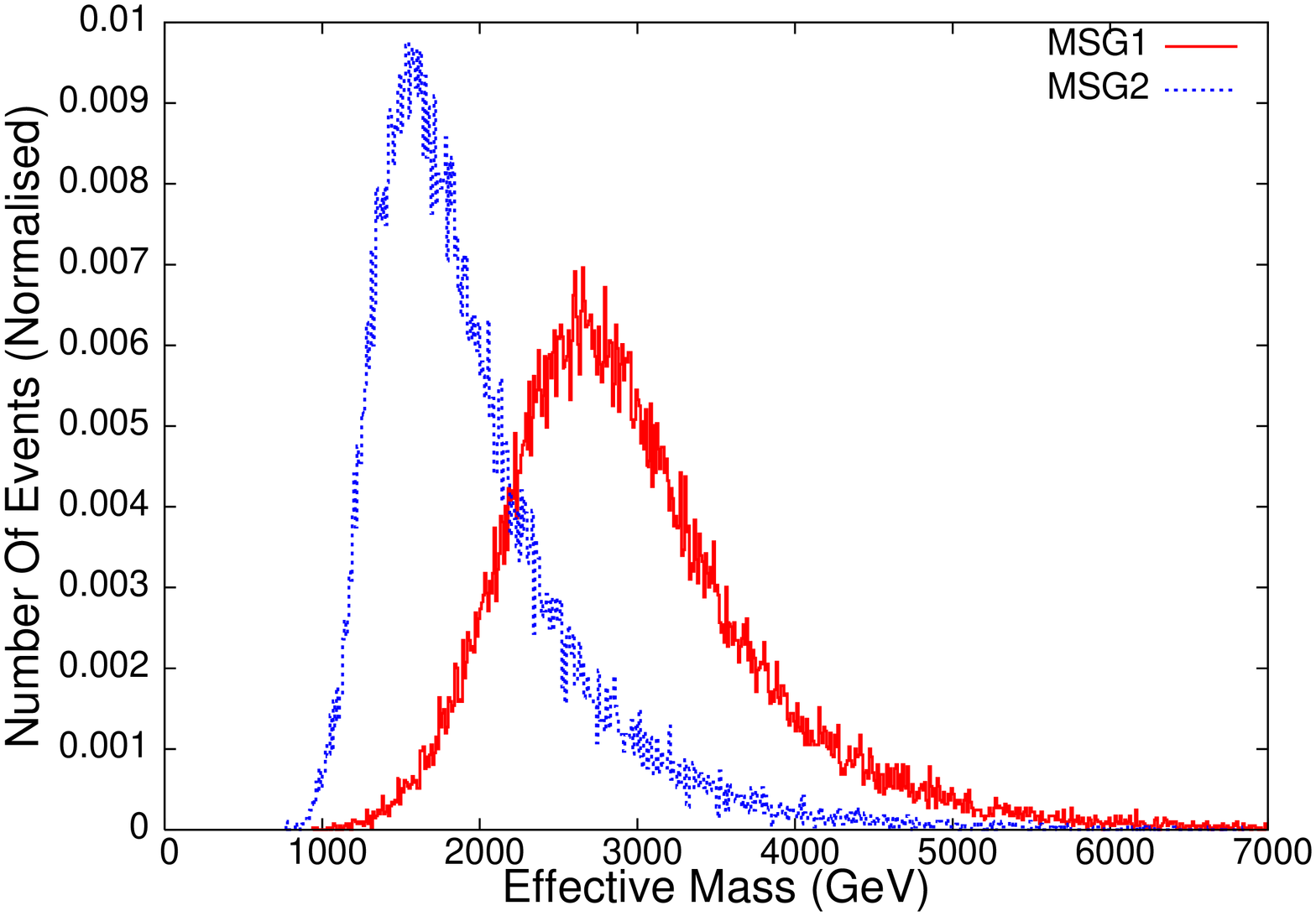,width=6.5 cm,height=7.5cm,angle=-90}}
\caption{{\em Effective mass distribution in $OSD$ events. 
{\tt Top left}: BP1 ({\em Red}), BP2 ({\em Blue}) and BP3 ({\em Pink}) chosen from {\em 770-422} with $\tan \beta$=5
have been plotted. {\tt Top right}: BP4 ({\em Red}) and BP5 ({\em Blue}) chosen from {\em 770-422} with $\tan \beta$=40
and $t\bar t$ ({\em Pink}) have been plotted. {\tt Bottom}: MSG1 ({\em Red}) and MSG2 ({\em Blue}) 
chosen from mSUGRA with $\tan \beta$=5 have been plotted. {\tt CTEQ5L} pdfset was used. 
Factorisation and Renormalisation scale has been set to $\mu_F=\mu_R=\sqrt{\hat s}$, 
sub-process centre of mass energy.}} 
\end{center}
\end{figure}

The missing energy distributions in OSD events at all the benchmark points have been shown in Figure 4. The 
organisation of the points remain the same as in Figure 3. In each case, the distribution starts from 100 GeV, 
as the event selection itself had this missing energy cut. As a result, all the points show a similar 
falling feature which indicates that the peak of the distribution is either small or around 100 GeV. 
The difference in the distributions is in the tail and is due to the hierarchy of the lightest neutralino masses. 
The heavier is the neutralino, the flatter is the distribution. Although this gives a nice distinction 
between the points with different gluino masses (and hence with different LSP masses), it is 
again, difficult to distinguish points with similar gluino masses.     

\begin{figure}[htbp]
\begin{center}
\centerline{\psfig{file=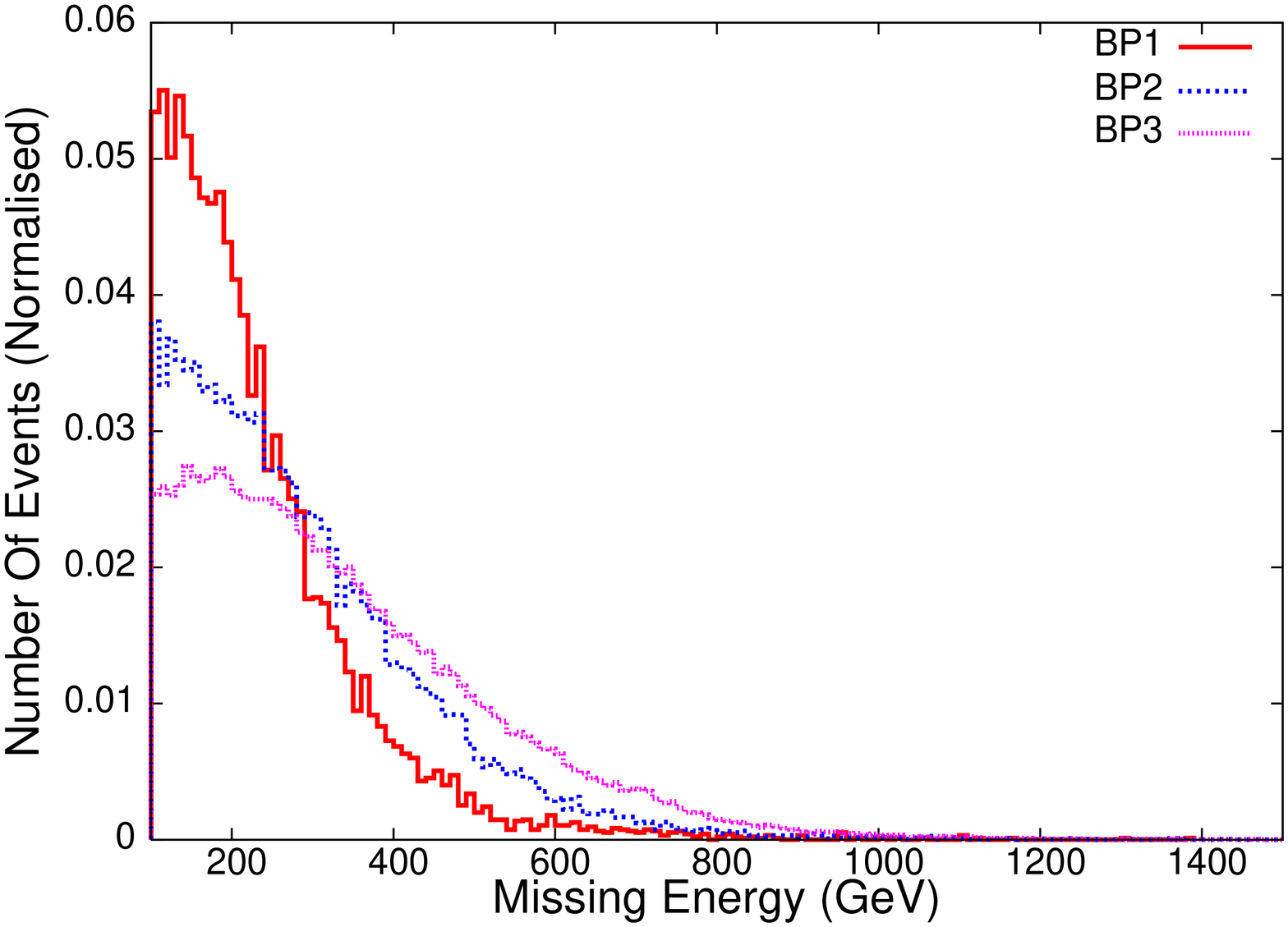,width=6.5 cm,height=7.5cm,angle=-90}
\hskip 20pt \psfig{file=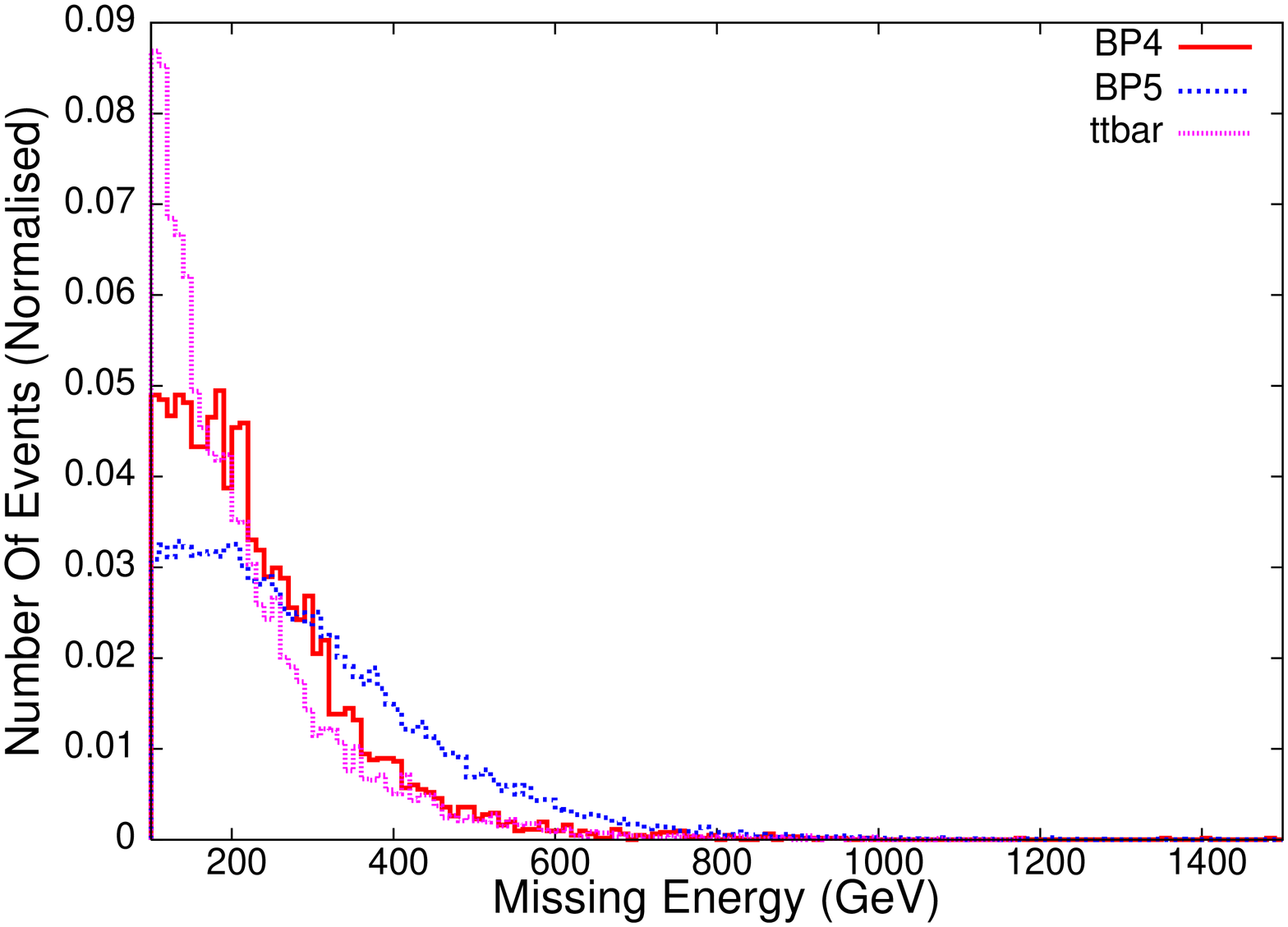,width=6.5 cm,height=7.5cm,angle=-90}}
\vskip 10pt
\centerline{\psfig{file=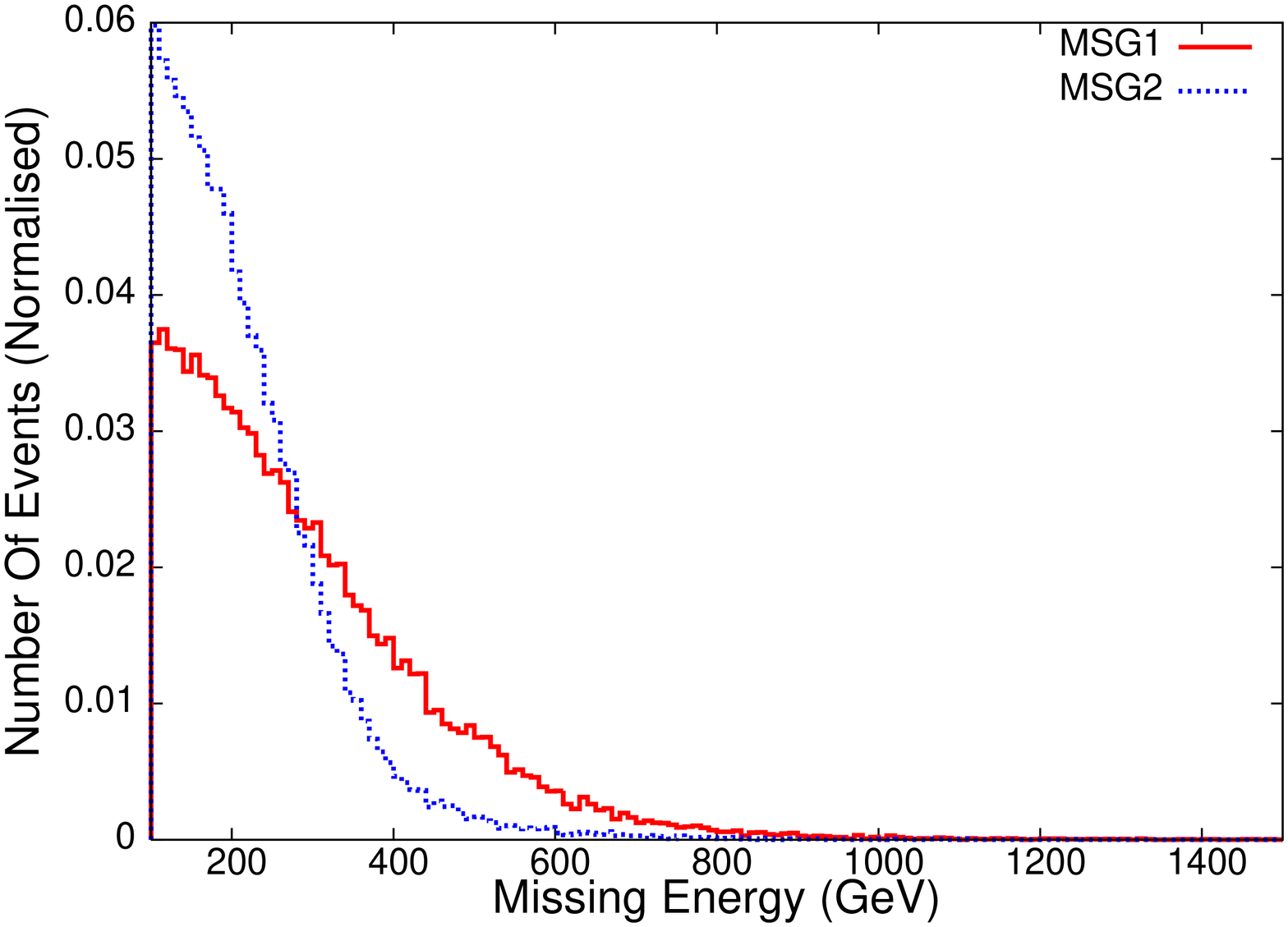,width=6.5 cm,height=7.5cm,angle=-90}}
\caption{{\em Missing energy distribution in $OSD$ events. 
{\tt Top left}: BP1 ({\em Red}), BP2 ({\em Blue}) and BP3 ({\em Pink}) chosen from {\em 770-422} with $\tan \beta$=5
have been plotted. {\tt Top right}: BP4 ({\em Red}) and BP5 ({\em Blue}) chosen from {\em 770-422} with $\tan \beta$=40
and $t\bar t$ ({\em Pink}) have been plotted. {\tt Bottom}: MSG1 ({\em Red}) and MSG2 ({\em Blue}) 
chosen from mSUGRA with $\tan \beta$=5 have been plotted. {\tt CTEQ5L} pdfset was used. Factorisation and 
Renormalisation scale has been set to $\mu_F=\mu_R=\sqrt{\hat s}$, sub-process centre of mass energy.}} 
\end{center}
\end{figure}

The numerical values of the event rates at the benchmark points 
are presented in Table 7, while Table 8 contains the results in similar 
channels for the mSUGRA ones. We note the contributions to these channels 
from the SM background in Table 9. While we note that the results are widely different 
from each other for different BPs, we also point out the distinction with 
corresponding mSUGRA ones with similar gluino masses. For example, when we
compare BP1, BP4 and MSG2 (all with gluino masses around 500 GeV), we note 
that the mSUGRA point yields much more events in almost all channels. This 
primarily has two reasons: one, the choice of the scalar mass parameter 
is very low for the mSUGRA one, compared to the non-universal case to obey CDM constraint
and two, the non-universal scenario studied here, have higher values of $M_1$ and 
$M_2$ at the high scale, which make the low-lying charginos and neutralinos heavier and 
correspondingly lower decay branching fraction through these to the leptonic final states.
Similar observation can be made in an attempt to compare the event rates of BP2, BP3, BP5 and
MSG1 ($m_{\tilde g}\simeq$ 1000 GeV). The reason that BP2 has slightly higher event rates 
than MSG1 can be attributed to the fact that $m_{\tilde g}$= 938.8 GeV for BP1, which is 
smaller to be compared to MSG1 ($m_{\tilde g}$= 1104 GeV).  

While we see that almost all the channels at the BPs rise sufficiently over the background 
fluctuations, the hadronically quiet trilepton channel gets submerged in to the background 
even for an integrated luminosity 100$fb^{-1}$ at the points BP2, BP3, BP5 and MSG1. 
This is because of the very high values of $m_{\tilde {\chi_2}^0}$ and $m_{\tilde {\chi_1}^{\pm}}$ with 
very small $\Delta m_{(\tilde {\chi_2}^0/\tilde {\chi_1}^{\pm}-\tilde {\chi_1}^0)}$. The significance 
in most of these points, in most of the channels (excepting the ${3\ell}$), are so high that it 
is very unlikely to be affected by the systematic errors. We would also like to point out that all 
the channels in BP1, BP4 and MSG2 rise over the background even for an integrated luminosity of 10$fb^{-1}$ 
(only exception being the $\sigma_{3\ell}$ for BP4), while others are suppressed by the background.
In Table 10, we summarise this information for each of the channels 
and parameter points, for an integrated luminosity of 30$fb^{-1}$. 

In the tables, the cross-sections are named as follows:
 $\sigma_{OSD}$ for OSD, $\sigma_{SSD}$ for SSD, $\sigma_{3\ell+jets}$ 
for $(3\ell+jets)$, $\sigma_{3\ell}$ for $(3\ell)$ and 
$\sigma_{4\ell}$ for inclusive 4 lepton events $4\ell$.
\vspace{0.4cm}

\noindent

\begin{center}

\begin{tabular}{|c|c|c|c|c|c|}

\hline
Benchmark Points & $\sigma_{OSD}$ & $\sigma_{SSD}$ &$\sigma_{3\ell+jets}$& 
$\sigma_{3\ell}$ & $\sigma_{4\ell}$ \\
\hline
\hline 
BP1 & 597.7 & 102.9 & 57.6 & 13.10 & 10.30 \\
\hline
BP2 & 70.1 & 15.7 & 11.6 & 0.09 & 1.64 \\
\hline
BP3 & 17.9 & 2.2 & 4.0 & 0.01 & 1.16 \\
\hline
BP4 & 398.4 & 120.8 & 30.8 & 4.50 & 3.28 \\
\hline
BP5 & 26.5 & 12.2 & 3.9 & 0.05 & 0.33 \\
\hline
\hline
\end {tabular}\\
\end{center}
\vspace{0.2cm}
{Table 7: {\em Event-rates (fb) in multilepton channels at the chosen benchmark points. 
{\tt CTEQ5L} pdfset was used. Factorisation and Renormalisation scale
has been set to $\mu_F=\mu_R=\sqrt{\hat s}$, subprocess centre of mass energy.}}
\noindent
\vspace{0.4cm}
\begin{center}

\begin{tabular}{|c|c|c|c|c|c|}

\hline
mSUGRA Points & $\sigma_{OSD}$ & $\sigma_{SSD}$ &$\sigma_{3\ell+jets}$& 
$\sigma_{3\ell}$ & $\sigma_{4\ell}$ \\
\hline
\hline 
MSG1 & 50.9 & 16.4 & 10.6 & 0.08 & 1.34 \\
\hline
MSG2 & 2781.8 & 175.9 & 285.1 & 12.26 & 99.89 \\
\hline
\hline
\end {tabular}\\
\end{center}
\vspace{0.2cm}
{Table 8: {\em Event-rates (fb) in multilepton channels at the mSUGRA points. 
{\tt CTEQ5L} pdfset was used. Factorisation and Renormalisation scale
has been set to $\mu_F=\mu_R=\sqrt{\hat s}$, subprocess centre of mass energy.}}
\vspace{0.4cm}

\begin{center}

\begin{tabular}{|c|c|c|c|c|c|}
\hline
\hline
SM Processes & $\sigma_{OSD}$ & $\sigma_{SSD}$ &$\sigma_{3\ell+jets}$& 
$\sigma_{3\ell}$ & $\sigma_{4\ell}$ \\
\hline
\hline
{\bf $t\bar t$} & 1102 & 18.1 & 2.7 & 5.3 & 0.0 \\
\hline
{\bf $ZZ$,$WZ$,$ZH$,$Z\gamma$} & 16.3 & 0.3 & 0.5 & 1.1 & 0.4 \\
\hline
{\bf Total} & 1118.3 & 18.4 & 3.2 & 6.4 & 0.4 \\
\hline
\hline
\end {tabular}\\
\end{center}
\vspace{0.2cm}
{Table 9: {\em Event-rates after cut (fb) in multilepton channels from the 
dominant SM backgrounds. The event rates in different channels for 
$t\bar t$ production have been multiplied by by proper $K$-factor (2.23) 
to obtain the usually noted NLO+NLL cross-section of $t\bar t$ \cite{ttbar}. 
{\tt CTEQ5L} pdfset was used. Factorisation and Renormalisation scale has been set to 
$\mu_F=\mu_R=\sqrt{\hat s}$, subprocess centre of mass energy.}}
\vspace{0.4cm}

\begin{center}

\begin{tabular}{|c|c|c|c|c|c|}

\hline
Model Points & ${OSD}$ & ${SSD}$ &${3\ell+jets}$& 
${3\ell}$ & ${4\ell}$ \\
\hline
\hline 
 {\bf BP1} & $\surd$ & $\surd$ & $\surd$ & $\surd$ & $\surd$ \\
\hline
 {\bf BP2} & $\surd$ & $\surd$ & $\surd$ & $\times$ & $\surd$ \\
\hline
 {\bf BP3} & $\times$ & $\times$ & $\surd$ & $\times$ & $\surd$  \\
\hline
 {\bf BP4} & $\surd$ & $\surd$ & $\surd$ & $\surd$ & $\surd$   \\
\hline
 {\bf BP5} & $\surd$ & $\surd$ & $\surd$ & $\times$ & $\times$ \\
\hline
 {\bf MSG1} & $\surd$ & $\surd$ & $\surd$ & $\times$ & $\surd$   \\
\hline
{\bf MSG2} & $\surd$ & $\surd$ & $\surd$ & $\surd$ & $\surd$   \\
\hline
\hline
\end {tabular}
\end{center}
{Table 10 : {\em $5 \sigma$ visibility of various signals for an integrated luminosity 
of $30 fb^{-1}$. A $\surd$ indicates a positive conclusion while a 
$\times$ indicates a negative one.}}
\vspace{0.4cm}

We also compare these results in the ratio space of events which is demonstrated in Figure 5
in form of barplot. The advantage of going to the ratio space is the uncertainties due to 
the choice of pdfsets, jet energy scale get reduced. Here we take the ratios of all events with 
respect to the OSD and referred as SSD/OSD, 3L+JETS/OSD, 3L/OSD and 4L/OSD along the x-axis of the 
barplot. As earlier, we divide the BPs and the MSGs in two categories: one, with BP1, BP4 and MSG2 
($m_{\tilde g}\simeq$ 500 GeV), which is shown in the left panel of the Figure 5; two, 
BP2, BP3, BP5 and MSG1 ($m_{\tilde g}\simeq$ 1000 GeV), which is shown in the right panel of the Figure 5.
We note that BP1, BP4 and MSG2 are well distinguished from SSD/OSD, 3L/OSD and 4L/OSD. While BP2
and MSG1 can not be distinguished very well from each other, identification of BP3, BP5 and MSG1 is quite
apparent from SSD/OSD and 4L/OSD events. 

In a nutshell we can summarise that, it is indeed possible to distinguish the non-universal 
gaugino mass scenario advocated here, from the mSUGRA ones with similar gluino masses. 
This is possible in both the absolute event rates or from the ratio plots shown here. However, 
the distinguishability reduces with increasing gluino mass.

\begin{figure}[htbp]
\begin{center}
\centerline{\epsfig{file=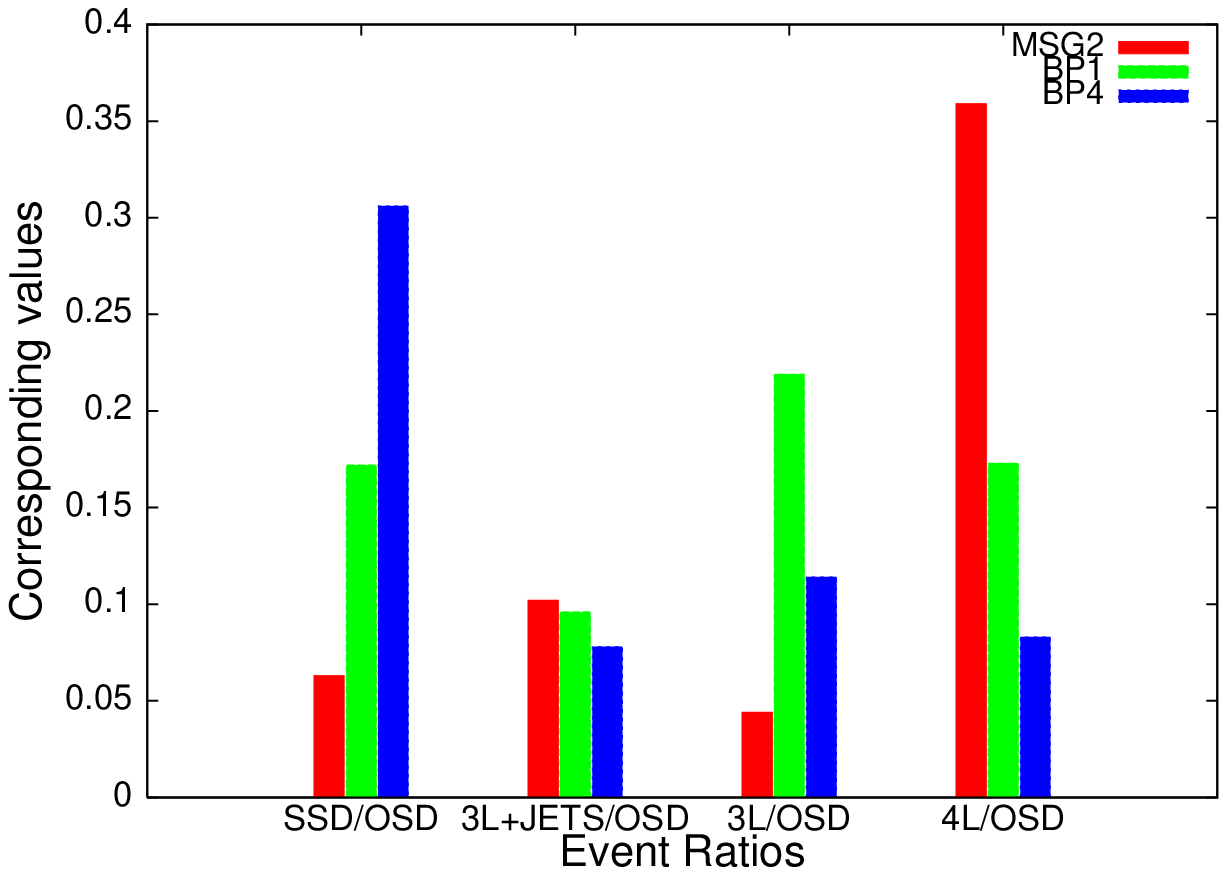,width=7.5 cm,height=6.5cm,angle=-0}
\hskip 20pt \epsfig{file=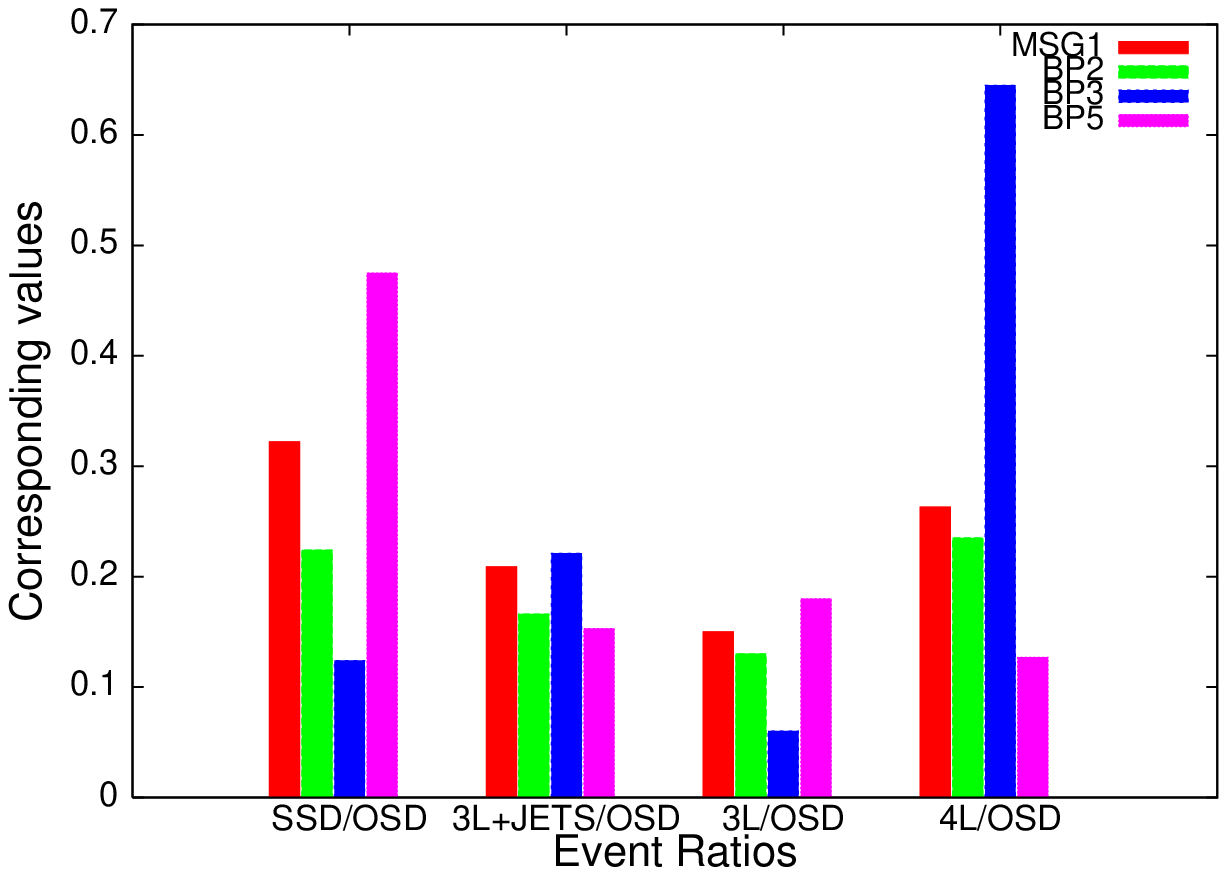,width=7.5 cm,height=6.5cm,angle=-0}}
\caption{{\em Event ratios of different benchmark points compared with mSUGRA.
Along the x-axis, ratios of events with respect to $OSD$ have been taken. The 
relative values of these ratios for different benchmark points are indicated 
along the y-axis. {\tt Figure on the left}: MSG2 ({\em Red}), BP1 ({\em Green}) and 
BP4 ({\em Blue}) are compared. $m_{\tilde g}$ is around 500 GeV in each case. 
{\em 3L/OSD} and {\em 4L/OSD} have been multiplied by 10. 
{\tt Figure on the right}: MSG1 ({\em Red}), BP2 ({\em Green}), BP3 ({\em Blue}) and BP5 
({\em Pink}) are compared. $m_{\tilde g}$ is around 1000 GeV in each case. 
{\em 3L/OSD} has been multiplied by 100 and 
{\em 4L/OSD} has been multiplied by 10 to accommodate them 
in the same figure. {\tt CTEQ5L} pdfset was used. Factorisation and 
Renormalisation scale has been set to $\mu_F=\mu_R=\sqrt{\hat s}$, 
sub-process centre of mass energy.}} 
\end{center}
\end{figure}

\section{Summary and Conclusions}

We have derived non-universal gaugino mass ratios for the representations 
{\bf 54} and {\bf 770} for the breaking chain 
$SU(4)_C \times SU(2)_L \times SU(2)_R~(G_{422})$ in a $SO(10)$ SUSY-GUT
scenario. We have assumed that the breaking of $SO(10)$ to the intermediate gauge
group and the latter in turn to the SM gauge group takes place at the GUT scale 
itself. We point out some errors in the earlier calculation and 
derive new results on the gaugino mass ratios. We scan the parameter space
with different low-energy constraints taken into account and point out the allowed 
region of the parameter space. We also study the dark matter constraint 
in these models and study collider simulation at some selected benchmark 
points in context of the LHC. The scans presented, have many interesting 
features that might help us in understanding the correlation between 
high scale input and low-energy spectra. We must mention here that the study 
is limited by the assumption that the series of symmetry breaking occurs at
the GUT scale itself. It is essentially a simplification, 
although we know that the mass relation in Equation (\ref {mr1}) is not 
going to change with different choices of the intermediate scale, 
while the running of the masses from the GUT scale to the intermediate 
scale will eventually change the gaugino mass ratios. However, our choice 
of the non-singlet Higgses, apart from conserving $D$-parity, are compatible 
even if the intermediate scale is different from the GUT scale.  
Within such a framework we have performed a collider study which is more 
illustrative than exhaustive. It nonetheless elicits some characteristics 
of the signature space for such high scale ratios in context of the LHC.
It is worth mentioning in this context that the comparison between the 
non-universal inputs with the mSUGRA ones yields
a significant distinction in the multilepton channel parameter space in context of the LHC.
This feature might be important in pointing out the departure of 
the 'signature space' with the inclusion of non-universality at the high scale.

While this paper was in preparation, we came across the reference \cite{nugmso10}, 
where the issue of gaugino mass non-universality in the context of 
$SO(10)$ and $E(6)$ has been addressed. While we agree completely with
the corrected gaugino mass ratios for $G_{422}$, our analysis in addition, 
point out the phenomenologically viable situations where
the choices of the Higgses get restricted with $D$-parity conservation and 
inclusion of the intermediate scale different from the GUT scale. 
Furthermore, the low-energy particle spectra in different cases have been 
derived in a comparative manner and the allowed regions of the parameter 
space consistent with low-energy and dark matter constraints are obtained in each case. 
In addition, we have predicted event-rates for such a breaking chain  
in a multichannel study pertinent to the LHC. The distinguishability of relative
rates in different channels has also been explicitly demonstrated by us.   

\vspace{0.2 cm}
\noindent
{\large {\bf Acknowledgment:}}
 
\vspace{0.2 cm}

We thank Biswarup Mukhopadhyaya and Amitava Raychaudhuri for many useful 
suggestions. We thank Sanjoy Biswas, Debottam Das and Nishita Desai for their 
technical help. We also acknowledge some very helpful comments from AseshKrishna Datta 
and Sourav Roy. This work was partially supported by funding available from the Department 
of Atomic Energy, Government of India for the 
Regional Centre for Accelerator-based Particle Physics, 
Harish-Chandra Research Institute. Computational work for this study was
partially carried out at the cluster computing facility of
Harish-Chandra Research Institute ({\tt http:/$\!$/cluster.mri.ernet.in}). 

\newpage
\noindent
{\Large{\bf Appendix}}\\

Using Equation (15) in Section 2, $T_Y$, $T_{B-L}$ and $T_{3R}$ in 45-dimension
can be written as:
\bea
T_Y=\sqrt{\frac{3}{5}}diag(\underbrace{0,..0}_8,\frac{2}{3},\frac{2}{3},\frac{2}{3},-\frac{2}{3},-\frac{2}{3},
     -\frac{2}{3},\underbrace{0,..0}_4,1,0,-1,\underbrace{\frac{1}{6},..,\frac{1}{6}}_6,
       \underbrace{-\frac{5}{6},..,-\frac{5}{6}}_6,\underbrace{\frac{5}{6},..,\frac{5}{6}}_6,
       \underbrace{-\frac{1}{6},..,-\frac{1}{6}}_6)
\eea

\bea
T_{B-L}=\sqrt{\frac{3}{8}}diag(\underbrace{0,..0}_8,\frac{4}{3},\frac{4}{3},\frac{4}{3},-\frac{4}{3},-\frac{4}{3},
     -\frac{4}{3},\underbrace{0,..0}_6,\underbrace{-\frac{2}{3},..,-\frac{2}{3}}_{12},
       \underbrace{\frac{2}{3},..,\frac{2}{3}}_{12})
\eea

\bea
T_{3R}=diag(\underbrace{0,..0}_{15},\underbrace{0,..0}_3,1,0,-1,
         \underbrace{\frac{1}{2},..,\frac{1}{2}}_6,\underbrace{-\frac{1}{2},..,-\frac{1}{2}}_6,
         \underbrace{\frac{1}{2},..,\frac{1}{2}}_6,\underbrace{-\frac{1}{2},..,-\frac{1}{2}}_6)
\eea
These lead to the following relation
\bea
T_{Y}=\sqrt{\frac{3}{5}} T_{3R}+\sqrt{\frac{2}{5}} T_{B-L}
\eea

\end{document}